# Fragmentation of spherical radioactive heavy nuclei as a novel probe of transient effects in fission

C. Schmitt,[1,2,*] K.-H. Schmidt,[1] A. Kelić,[1] A. Heinz,[3] B. Jurado,[4] P.N. Nadtochy[5]

[1] *Gesellschaft fuer Schwerionenforschung, 64291 Darmstadt, Germany*
[2] *Grand Accélérateur National d'Ions Lourds, BP 55027, 14076 Caen Cedex 05, France*
[3] *Wright Nuclear Structure Laboratory, Yale University, New Haven, CT 06520, United States*
[4] *Université Bordeaux I, CNRS/IN2P3, CENBG, BP 120, 33175 Gradignan, France*
[5] *Omsk State University, Department of Theoretical Physics, 644077 Omsk, Russia*

## *Abstract*

Peripheral collisions with radioactive heavy-ion beams at relativistic energies are discussed as an innovative approach for probing the transient regime experienced by fissile systems evolving towards quasi-equilibrium, and thereby studying the viscous nature of nuclear matter. A dedicated experiment using the advanced technical installations of GSI, Darmstadt, permitted to realize ideal conditions for the investigation of relaxation effects in the meta-stable well. Combined with a highly sensitive experimental signature, it provides a measure of the transient effects with respect to the flux over the fission barrier. Within a two-step reaction process, 45 proton-rich unstable spherical isotopes between At and Th produced by projectile-fragmentation of a stable $^{238}$U beam have been used as secondary projectiles which impinge on lead target nuclei. The fragmentation of the radioactive projectiles results in nearly spherical compound nuclei which span a wide range in excitation energy and fissility. The decay of these excited systems by fission is studied with a dedicated set-up which, together with the inverse kinematics of the reaction, permits the detection of both fission products in coincidence and the determination of their atomic numbers with high resolution. The information on the nuclear charges of the two fragments is used to sort the data according to the initial excitation energy and fissility of the compound nucleus. The width of the fission-fragment nuclear charge distribution is shown to be specifically sensitive to pre-saddle transient effects and is used to establish a clock for the passage of the saddle point. The comparison of the experimental results with model calculations points to a fission delay $\tau_{trans}$ of $(3.3\pm0.7)\cdot10^{-21}$s for initially spherical compound nuclei, independent of excitation energy and fissility. This value suggests a nuclear dissipation strength $\beta$ at small deformation of $(4.5\pm0.5)\cdot10^{21}$s$^{-1}$. The very specific combination of the physics and technical equipment exploited in this work sheds light on previous controversial conclusions, which were drawn without considering the influence of the initial conditions.

## I. Introduction

The escape from a meta-stable state is a problem which appears in many fields as various as fluid mechanics, chemistry or nuclear physics [1]. Systems which are initially out of equilibrium display a transient behaviour [2, 3] during which relaxation in all degrees of freedom takes place. Once this is achieved, a quasi-equilibrium regime is reached. The finite time required to establish quasi-equilibrium results in a complex time-dependent behaviour of the decay rate of the meta-stable state. Transient effects have no perceptible influence on the decay of the system, if the lifetime of the meta-stable state is much longer than the relaxation time of the typical degrees of freedom. In this case, quasi-equilibrium can be assumed as established immediately. This is the basic idea behind the transition-state theory [4]: Passage over the barrier is exclusively governed by the available phase space. Only if the characteristic relaxation times make up a sizeable fraction of the lifetime of the system, the transient regime may manifest itself. Direct observation of transient effects relies on the availability of a fast clock, suited for measuring the decay time of the system. This defines the conditions needed for transients to become evident.

---

[*] Electronic address : schmitt@ganil.fr

The understanding of transient effects is important due to its direct connection with the viscosity of the medium. A purely microscopic description of nuclear dissipation is presently a challenge [5, 6], and it is customary to describe the decay of the meta-stable state by using hybrid models, namely transport theories [7]. The latter distinguish between collective and intrinsic degrees of freedom. Intrinsic excitations are based on the states of the individual constituents and form a heat bath. Collective modes correspond to a coordinated motion involving most of the constituents. The projection of the many-body problem onto a macroscopic scale restricted to the most relevant coordinates introduces the notion of coupling, i.e. energy exchange between the collective degrees of freedom and the heat bath. Friction retards the collective motion due to the irreversible energy flow from collective modes to intrinsic excitations. It is a good approximation for many applications in everyday life. In the nuclear medium, depending on the initial conditions, additional diffusion processes which lead to an exchange of energy in both directions can not be neglected. Here, the dynamical evolution of the system is given by the solution of either the Fokker-Planck or the Langevin classical equation of motion [7, 8], where the combined action of the driving potential, friction and diffusion forces is computed. The differential Fokker-Planck equation gives access to the distribution probability of the collective degrees of freedom as a function of time along the fission path. The integral Langevin equation traces the time evolution of the system step by step for individual trajectories. The solution of the equation of motion shows that dissipative friction and diffusion phenomena play an important role in the decay of the meta-stable state and, namely, that they affect the duration of the transient regime [2]. Additionally, the transitional saddle is found not being an absorption point as assumed in the transition-state theory [4]: Fluctuations across the barrier can lead to multiple passages over the saddle point before the decay is decided [9, 10]. The actual quasi-stationary escape rate is thus reduced as compared to the transition-state value by a factor $K$ which was originally determined by Kramers [11]. Transport theories have proven relevant in various fields [12] among which nuclear physics is one [8]. In this context, collective modes correspond to vibration, elongation, constriction, etc, while the heat bath is based on nucleonic single-particle states.

As a typical large-scale amplitude motion involving a meta-stable state, fission is an excellent probe of pre-saddle relaxation effects at the subatomic level. The product of a given reaction mechanism is, in general, out of equilibrium. The collective coordinates, which govern the motion in the fission direction, have generally a larger relaxation time than the intrinsic degrees of freedom. That is, collective modes need a finite time for adjusting to the potential-energy landscape and establishing a quasi-stationary flux across the fission barrier. A transient regime seems therefore natural and inevitable. The potential forces drive the mean value of the probability distribution to the minimum of the potential well while dissipative forces enlarge its width. Due to this broadening, all states in the quasi-bound region, including those above the fission saddle, are progressively populated in accordance with the available phase space, and the system can eventually escape from the meta-stable region. The complex influence of the friction and diffusion forces in the quasi-bound region leads to a fission decay-width $\Gamma_f(t)$ which explicitly depends on time. As discussed above, before the quasi-stationary regime is reached, fission is fully or partly inhibited. This delay of fission is commonly called the transient time $\tau_{trans}$ [2, 3][1]. Beyond the saddle, Coulomb repulsion drives the nucleus to more elongated shapes until it splits into two fragments and dissipative phenomena are dominated by friction which increases the saddle-to-scission time $\tau_{ss}$ [13, 14, 15].

Although the very nature of nuclear dissipation remains an open question [16, 17, 18, 19, 20], its influence on the total fission time is well-established [21]. In contrast, the existence of pre-saddle relaxation effects, its magnitude and potential manifestation are debated [22, 23, 24]. This question has an important impact on our fundamental understanding of nuclear dynamics and viscosity. In addition, there are experimental indications [21] and theoretical expectations (see e.g. [16, 17]) that the dissipation strength depends on the deformation of the system. This calls for the study of the dissipation at small (inside the saddle point) and large (between saddle and scission) deformations independently. Discriminating pre- and post-saddle effects might be the only way for unambiguously probing dissipative phenomena in fission [25]. Previous investigations on dissipation at small deformation are scarce and the conclusions are contradictory. In the present work, very specific favourable conditions for studying nuclear transient effects and dissipation over a small deformation range around the spherical shape were reached for the first time, making use of advanced technical installations, highly sensitive experimental signatures and elaborate model calculations.

Aside from fundamental interest, a better understanding of fission dynamics is crucial with respect to other nuclear science aspects like, astrophysics [26], super-heavy element synthesis [27] and super- or hyper-deformation in nuclei [28]. Also, important applications concern nuclide production at radioactive beam facilities [29], accelerator driven reactor systems [30], nuclear power installations, safety and security [31, 32, 33].

**1) Initial conditions and dynamics**

From the very definition, an ideal scenario for evidencing transient effects consists of starting with a system whose collective ("slow") degrees of freedom are initially out-of-equilibrium, while all other ("fast") degrees of freedom are equilibrated. A simple test-case was investigated by Grangé and collaborators [2, 3]. They showed that ideal initial conditions for probing the growth of the flux at the fission saddle point are met with initially highly excited and fissile spherical nuclei [34]. When starting from the bottom of a spherical potential well, transient effects are expected to be effective [35] and their manifestation should be large. Furthermore, because the probability distribution is initially localized at the spherical minimum of the well, driving forces are absent, and the shape evolution is exclusively governed by diffusion. In contrast, the descent from saddle to scission is dominated by the friction force. Thus, in this ideal scenario, the two stages of the fission process allow probing selectively friction and diffusion in nuclei [36]. Preparing a system which fulfils these theoretical initial conditions is an experimental challenge. The nucleonic degrees of freedom of a cold nucleus, which is spherical in its ground state, have to be suddenly excited without affecting the other (collective) degrees of freedom. In the present work, we report on the first experimental realization of the ideal scenario of a highly fissile nucleus with a nearly spherical shape, high excitation energies and low angular momenta. The limited deformation range probed with this ideal scenario is further particularly well suited for extracting information on the dissipation strength $\beta$ at small deformation. Interference with effects related to the potentially strong variation of $\beta$ with deformation [37, 38] remains weak.

---

[1] The transient time was defined by Grangé and collaborators [2, 3] as the time required for the fission decay-width to reach 90% of its asymptotic value at quasi-equilibrium.

## 2) Experimental signature

Measuring a nuclear time is a difficult task. Direct measurements are based on crystal blocking [39] and electronic K-vacancy lifetime [40] techniques. Unfortunately, the sensitivity of these methods does not extend below $10^{-19}$-$10^{-18}$s, and only the tail of the total fission time distribution is accessible [41]. The most direct and efficient way to probe fission time scales is based on the number of light particles or γ-rays emitted by the system prior to scission. The longer it takes to the nucleus for reaching the scission point, the larger is the time that is left for de-excitation by particle evaporation and γ-ray emission. While the pre- and post-scission particles are discriminated by their kinematics, disentangling the pre- and post-scission photons de-exciting a Giant Dipole Resonance (GDR) is achieved according to the energy of the GDR peak which depends on the mass of the emitter. The concept of the particle [42, 43] and γ-ray [44] clock led to tremendous progress in our understanding of fission dynamics in the last decades. The number of pre-scission particles $M^{sci}$ gives, by definition, access to the total fission time. The study of transient effects requires isolating the pre-saddle time which is a part of the total fission time, only. This is very challenging since $\tau_{trans}$ is expected to be a few $10^{-21}$s [2, 3]. Yet, light-particle (namely neutron) emission times can be very small (of the order of $10^{-22}$s to $10^{-21}$s) at high excitation energies of heavy nuclei [45], and times as short as $\tau_{trans}$ can in principle be probed. Unfortunately, disentangling the pre- and post-saddle contributions to $M^{sci}$, and obtaining thereby insight into early transient effects, is experimentally impossible. It is thus customary to extract $\tau_{trans}$ with the help of model calculations (see e.g. Ref. [46]) from more indirect signatures.

The multiplicity of the emitted light particles or high-energy photons is inevitably related to the excitation energy of the decaying system. Aside from counting particles, an alternative relevant measure of pre-saddle evaporation is the measurement of the excitation energy at the saddle point: The longer the pre-saddle delay, the larger the probability of evaporating particles, which results in a lower excitation energy, and thus temperature $T^{sad}$, at saddle. Because the decision for fission is essentially made at the outer barrier, the fission probability depends on the modification of the excitation energy of the nucleus during $\tau_{trans}$. It seems therefore natural to exploit the fission and evaporation residue cross sections, $\sigma_f$ and $\sigma_{ER}$, respectively, to obtain insight into pre-saddle dynamics [47]. Nonetheless, the simultaneous modification of the excitation energy, angular momentum and fissility of the system caused by evaporation influences the decay probability of the remaining system in a complex way. As a consequence, attempts to combine pre-scission multiplicity and cross section data proved insufficient to constrain critical model parameters (see e.g. [48, 49, 50, 51, 52]). Angular momentum distributions, which are influenced by pre-saddle evaporation, depend also rather strongly on uncertain parameters [53, 54].

Observables connected more exclusively to $T^{sad}$ were therefore investigated. It was suggested to extract the temperature at the saddle point by measuring the angular distribution of the fission fragments whose anisotropy is proportional to $T^{sad}$ according to the Statistical Saddle Point Model [55]. However, the theoretical description remains difficult due to the influence of target deformation [56] and nuclear shell structure [57], as well as due to contributions from quasi-fission [58] and pre-equilibrium fission [59]. Furthermore, most studies are based on fusion-induced fission for which the initial angular momentum can be very large which brings additional complications.

An alternative pertinent measure of the temperature at saddle is the width $\sigma_A$ of the fission-fragment nuclear mass $A$ distribution. Experimental data [60, 61] suggest that the decision on mass partition is likely made quite early on the path to scission. Within the statistical model [62, 63] the width of the fission-fragment $A$ distribution can be parameterised as follows:

$$\sigma_A^2 = A_{fiss}^2 \cdot T^{sad} / (16 \cdot d^2V/d\eta^2) \quad (Eq.1)$$

with $d^2V/d\eta^2$ being the second derivative of the potential (stiffness hereafter) with respect to mass-asymmetric deformation $\eta = (4/A_{fiss})(M-A_{fiss}/2)$ at the saddle point. $A_{fiss}$ corresponds to the mass of the fissioning nucleus and $M$ to that of the fragment. The evaporation of particles by the compound system between saddle and scission and by the excited fragments after scission can partly smear out the close connection between $\sigma_A^2$ and $T^{sad}$. Due to the strong correlation which exists between nuclear mass and charge in fission, the width of the fission-fragment $Z$ distribution depends on $T^{sad}$ accordingly, and we have:

$$\sigma_Z^2 \approx (Z_{fiss}/A_{fiss})^2 \cdot \sigma_A^2 = Z_{fiss}^2 \cdot T^{sad} / (16 \cdot d^2V/d\eta^2) \quad (Eq.2)$$

where $Z_{fiss}$ is the nuclear charge of the fissioning nucleus. For heavy nuclei, neutron evaporation dominates over light-charged particle emission [64, 65, 66]. Consequently, the $Z$-width is expected to be less affected by evaporation than the $A$-width. This stronger correlation between $\sigma_Z$ and $T^{sad}$ makes the $Z$-width being especially sensitive to the pre-saddle dynamics[2]. The stiffness $d^2V/d\eta^2$ is a static quantity which has to be defined at a given deformation point along the fission trajectory. Here, the decisive influence on the fission-fragment mass (or equivalently, nuclear charge) distribution is attributed to the saddle point. It might be objected that this assumption does not include dynamical effects along the descent between saddle and scission which could influence the fission partition. This question has been debated for years. Recent dynamical calculations [67] based on the three-dimensional Langevin equation investigated the dependence of $\sigma_A^2$ on the angular momentum. It was noticed that, for heavy nuclei, the values obtained for $\sigma_A^2$ from the dynamical calculation for a given $L$ largely exceed the results of the statistical model when (Eq.1) is evaluated at the scission point. This was interpreted as a clear evidence for memory effects: According to Ryabov *et al.* [67], the nucleus "remembers the former large fluctuations of mass-asymmetry coordinate during descent from the saddle to scission". In other words, lower values of the stiffness, characteristics for the saddle point, seem to enter into play. The decision on mass and nuclear-charge partition is thus likely to be made quite early on the descent towards scission i.e. close to the outer saddle. From a comprehensive compilation of experimental fission-fragment mass distribution widths with initial excitation energies $E^*$ up to about 150 MeV, Rusanov *et al.* [60] parameterised $d^2V/d\eta^2$ at the saddle point as a function of fissility. This empirical parameterisation can be used together with (Eq.1) ((Eq.2)) to extract the temperature at the saddle from the measurement of $\sigma_A$ ($\sigma_Z$). The close correspondence between $\sigma_Z$ and $\tau_{trans}$ was exploited and demonstrated successful by Jurado *et al.* [68] up to very high excitation energies. Times as short as a few $10^{-22}$s are in principle within reach using the $\sigma_Z$ signature. The $Z$-width constitutes a reliable and robust signature of transient effects because of its transparent modelling at high excitation energies [60, 61, 68] and its weak dependence on model parameters (see [69] and section VI).

The present work investigates fission transient effects in very favourable conditions based on the above developed ideas. The study focuses on highly excited fissile systems, which are prepared in a spherical shape: Thus, transients are expected to be strong. As $\sigma_A$ can experimentally only be evaluated after post-scission evaporation, the width $\sigma_Z$, which defines $\sigma_A$ at the time the system passes the saddle point, is proposed to be used as a less ambiguous measure of the temperature at saddle. In Section II the innovative approach based on the fragmentation-induced fission of highly fissile spherical radioactive projectiles is presented. The coincident detection of the two fission fragments is shown as crucial for exploiting the rich information contained in the $\sigma_Z$ observable, allowing a classification of the data according to initial excitation energy and fissility. Section III is devoted to the experimental set-up. The general trend in the measured fission-fragment *Z* distributions is examined in Section IV. The manifestation of a finite fission delay at high temperature is demonstrated, and an accurate survey of the data, in connection with model calculations, is performed in Section V. The values for the transient time characteristic of initially spherical nuclei and for the dissipation strength at small deformation are extracted. The results are compared to previous works in Section VI, and some longstanding points of discord are discussed. A summary and concluding remarks are given in Section VII.

---

[2] Here we discuss the projection of the complete fission-fragment distribution in $Z$. This is different from the isobaric $Z$ distribution at fixed mass. The $N/Z$ degree of freedom, relevant for the latter, is known to evolve fast and it is determined close to the scission configuration [J.P.Bocquet *et al.*, Zeit. für Phys. A 335 (1990) 41].

## II. The fragmentation-induced fission scenario

The nuclear system produced in a collision is characterized by its initial distributions in excitation energy $E^*$, angular momentum *L* and deformation as well as by its mass and nuclear charge, all of which depend on the entrance-channel reaction mechanism. Achieving the aforementioned ideal scenario was so far hampered by the inability of producing highly excited and highly fissile spherical compound nuclei.

The advent of heavy-ion beams made it possible to form fissioning nuclei by fusion. An appropriate choice of the reaction partners and bombarding energy renders fission highly competitive with particle evaporation. While the excitation energy $E^*$ of the compound nucleus (CN) hardly exceeds 150 MeV, large angular momenta which increase the fission probability are accessible. The main difficulty inherent to fusion-induced reactions is that the distribution in $E^*$, *L* and deformation of the CN can be very broad [70, 53, 35]. Resolving the interplay between all these effects is difficult. A higher selectivity in the initial temperature and angular momentum is achievable with fission induced by nucleon transfer reactions. The excitation energy induced in these reactions, typically below 80 MeV, is nevertheless at the limit at which transients are expected to manifest efficiently [71], and the remaining structural effects can blur the interpretation [72]. In addition, the amount of highly fissile systems accessible is limited with this method. Finally, the required heavy fissile reaction partner is necessarily deformed, since $^{209}$Bi is the heaviest spherical primordial nucleus. Like in fusion-induced reactions, the initial system, although thermalized, can thus be sizeably deformed when the motion into fission direction starts [35]. Shorter transient delays are expected for such systems which start on the average closer to the saddle point. No realistic analytical modelling of the influence of the initial deformation exists at present. The omission of this aspect of the reaction in standard analysis may lead to erroneous conclusions [35]. Namely, it was shown that the dissipation strength might be underestimated if the influence of initial deformation of the CN is important and neglected [73].

Recently, peripheral collisions involving heavy projectiles at relativistic energies [74, 68] and spallation reactions (see Refs. [75, 76, 77, 78, 22] and therein) have proven to be a pertinent tool for the investigation of fission dynamics. The main advantage of these mechanisms relies on the properties of the produced compound nucleus. In a fragmentation reaction, the fast interaction [79, 80] between the projectile and the target leads to a remnant nucleus whose shape is nearly undistorted with respect to the shape of the projectile. In spallation, the cascade of collisions initiated by the light particle does not sizeably alter the shape of the heavy partner either. Collective excitations are small and the entity formed, hereafter called the *pre-fragment*, is rather well defined [81, 82]. In other words, the pre-fragment, which is the compound nucleus in this type of reactions, is characterized by a narrow shape distribution very similar to the initial configuration of the projectile (target) in inverse (normal) kinematics. Such a feature is particularly well suited for the study of transient effects, because the collective coordinates of the system are initially *not* equilibrated. Additionally, the intrinsic energy $E^*$ induced in the CN pre-fragment reaches up to high values [80, 83, 84, 85] for which transient effects are expected to manifest strongly due to highly probable particle evaporation during $\tau_{trans}$. Compared to spallation for which large angular momenta (up to 50 ℏ) can be reached in the pre-fragment [86], fragmentation induces only small angular momenta (10 ℏ on average) [87, 88, 89]. That limits the complexity of the analysis and the interpretation in the latter case [67]. The assets of fragmentation and spallation mechanisms were exploited to investigate pre-saddle transient effects in Refs. [68, 76, 77] using the stable $^{238}$U nucleus as a beam or a target. A major unknown in these studies is the influence of the initial deformation [35] of the pre-fragment due to the well-deformed character of $^{238}$U [90]. Since $^{238}$U and $^{232}$Th are the only fissile projectiles available, one is unavoidably left with initially deformed CN candidates. Investigations based on initially spherical systems are accessible with stable but lighter ions [75, 77]. They are limited in accuracy due to low fission probabilities. Furthermore, their sensitivity to nuclear dissipation is debated [91, 92, 93].

The challenge of matching the aforementioned theoretical scenario was overcome by using the state-of-the-art technical installations available at GSI, Darmstadt. The properties of proton-deficient actinides were exploited for producing highly excited and fissile compound nuclei with a rather well-defined initial shape configuration. An inspection of the nuclear chart reveals that proton-rich nuclei near $N = 126$ are the only systems which present both a spherical shape in their ground state and high fissility. However, these nuclei are not stable and can thus only be reached as radioactive projectiles. In praxis, producing the corresponding isotones is just possible at an in-flight secondary-beam facility based on fragmentation. The approach pioneered at GSI is based on the two-step mechanism [34, 94] illustrated in Fig.1, upper part. Fragmentation of a primary $^{238}$U beam produces a large variety of nuclei. These projectile-like products adapt to their ground-state configuration well before leaving the production target. Among them, proton-rich isotopes of elements between At and Th, with neutron numbers around $N = 126$, are selected. These nuclei, still at relativistic energies, act as secondary projectiles which impinge on a lead target. As outlined above, nuclear-induced fragmentation of fissile spherical projectiles leads to fissile and nearly undistorted pre-fragments at high $E^*$ and small $L$. Within the fast cut-off abrasion scenario [79, 80], we estimated the quadrupole deformation parameter $\beta_2$ induced in fragmentation of a heavy spherical projectile [95]. Figure 2 shows $\beta_2$ as a function of the pre-fragment mass $A_{prf}$. The average initial deformation of the nuclei which end in fission, and whose calculated abundances are displayed in the inset, is small and oblate with $|\beta_2| < 0.15$. This value can be compared to the significantly larger saddle-point deformation of $\beta_2 \approx (0.6\text{-}0.8)$ typical for heavy systems. Within the present fragmentation-fission approach, it is the transient regime of the pre-fragment formed in the abrasion of the secondary projectile which is searched for. The highly excited and fissile nearly spherical pre-fragment decays by a competition between fission and evaporation.

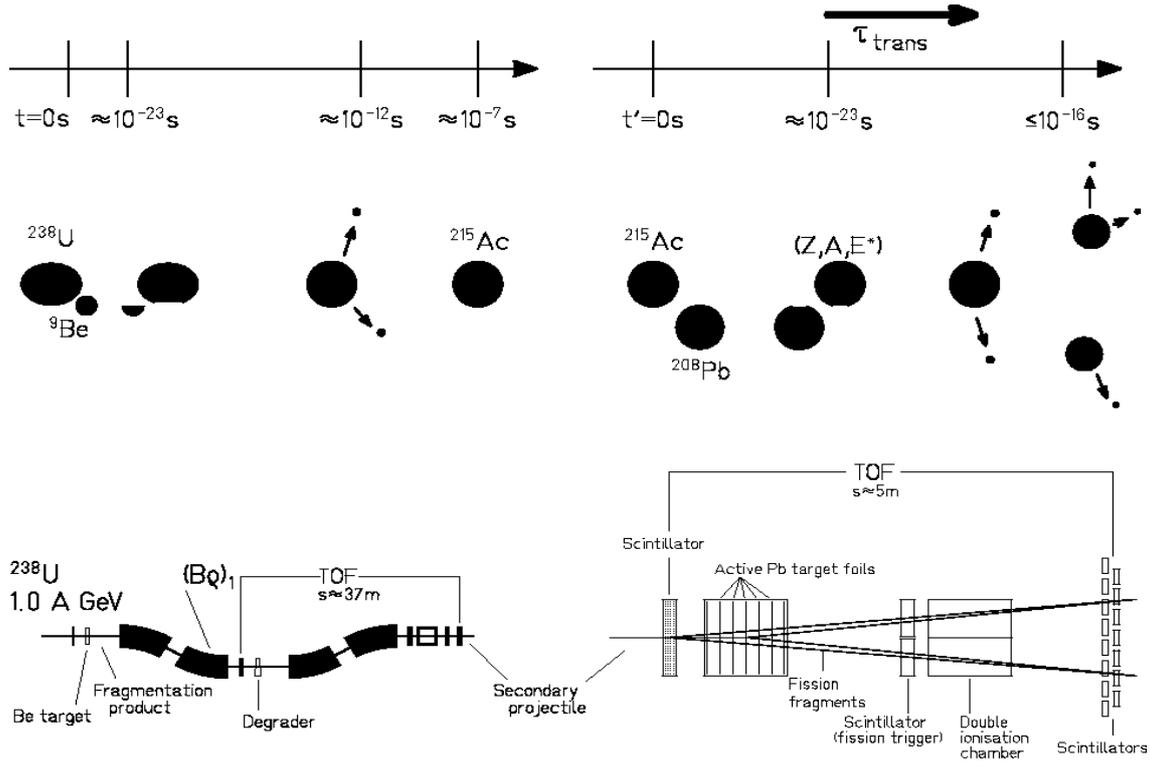

**Figure 1**: Top: Schematic drawing of the radioactive-beam production (left) and the subsequent fragmentation-induced fission reaction (right). Typical time scales t and t' referring to the primary and secondary reaction, respectively, are indicated. Bottom: Experimental set-up, consisting of the Fragment Separator (left) and the set-up for the detection of fission (right) located directly behind the separator. For sake of clarity, the scale used in the left and right drawings is different.

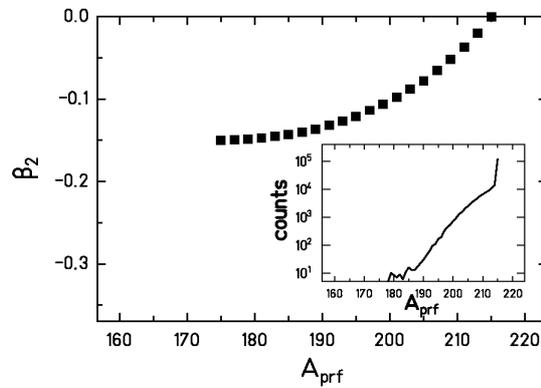

**Figure 2**: Calculated quadrupole deformation parameter $\beta_2$ of the pre-fragments formed in the abrasion of a spherical projectile of mass $A_{proj}$ =215 as a function of the pre-fragment mass $A_{prf}$. The inset shows the calculated $A_{prf}$ distribution of the pre-fragments originating from fragmentation of $^{215}$Ac on lead which decay by fission. The peak at $A_{prf}$ = 215 is due to fission induced by electromagnetic interactions.

Besides the fundamental quest for dissipative phenomena, the present work highlights the relevance of using fragmentation-induced reactions for studying nuclear dynamics. Hundreds of different nuclei are formed in a given reaction and a particularly wide excitation energy domain is spanned. The angular momentum and the deformation of the CN pre-fragment are quite well defined, minimizing complex side effects.

## III. Experimental set-up

In line with the reaction scenario, the experimental set-up used for the present measurement consists of two stages. A schematic view is given in Fig.1, lower panel. We restrict ourselves to the main features of the experiment and refer to [94] for details.

The fragmentation of a primary $^{238}$U beam at 1000 $A$MeV delivered by the heavy-ion synchrotron SIS onto a 657 mg/cm$^2$ thick beryllium target located at the entrance of the GSI fragment separator FRS [96], produces a large number of proton-rich nuclei. The FRS, with its ability to spatially separate and identify particles event-by-event, was used to prepare secondary beams from these fragmentation products of $^{238}$U. At the exit of the separator around 45 nearly spherical proton-rich isotopes of At up to Th at about 420 $A$MeV were available. With an average primary-beam intensity of $10^7$ particles per second, intensities of 100 products per second for a specific radioactive isotope could be obtained in favourable cases. The radioactive beams entered the second stage of the set-up, which was dedicated to the in-flight detection of the two fragments produced in the fragmentation-induced fission of the secondary projectile after its interaction in an active lead target. The latter consisted of five foils with a total thickness of 3.03 g/cm$^2$ mounted inside a gas-filled detector. They acted as a subdivided ionization chamber. This arrangement permitted to determine whether fission took place in either the target or the scintillator located in front of it. It aided in discriminating nuclear- from electromagnetic-induced reactions as discussed in Refs. [94, 97]. The inverse kinematics of the reaction leads to forward-focused fission fragments with high kinetic energies. This makes it possible to detect the two fragments in coincidence with nearly full solid-angle coverage, and to unambiguously identify them in atomic number $Z$ by the use of a double ionization chamber. The data are corrected for secondary reactions of the fission fragments on their way to the chamber according to the method developed in Ref. [98]. The achieved resolution amounts to $\Delta Z = 0.4$ (FWHM). The correlation spectrum displayed in Fig.3 demonstrates the excellent nuclear charge identification - each spot corresponds to one pair of fission fragments. The mass of the fission fragments could not be determined in this experiment. As outlined in section I the information on the element number is appropriate for the purpose of the present work.

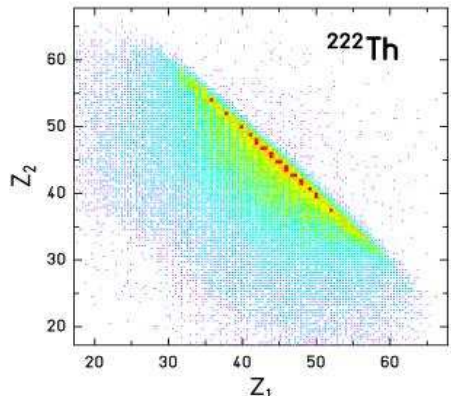

**Figure 3**: (Colour online) Experimental correlation between the nuclear charges $Z_1$ and $Z_2$ of the two fragments resulting from fragmentation-induced fission of radioactive $^{222}$Th projectiles impinging on lead nuclei. The displayed colour scales logarithmically with the observed intensity.

## IV. Analysis and results

## 1) Data analysis

Depending on the impact parameter, peripheral collisions with relativistic heavy ions proceed *via* electromagnetic or nuclear interactions. While the former collisions mainly excite giant resonances with $E^*$ hardly exceeding 20 MeV, the latter are more violent. The interference between Coulomb and nuclear processes was observed to be negligible [99], and the total fission cross section can be approximated by the sum of the independent electromagnetic and nuclear contributions. In the following we show how the admixture of events from electromagnetic- and nuclear-induced fragmentation-fission manifests itself in the measured $Z$ distributions. The ability of experimental filters, based on the coincident measurement of the two fragments, to classify the data according to the excitation energy and to the fissility of the pre-fragment is demonstrated.

### a. Contributions to fission induced by a fragmentation reaction

The fragmentation of a heavy projectile leads to a large amount of pre-fragments which differ substantially in their nuclear composition and excitation energy. The integral of the fission-fragment nuclear-charge cross sections measured in the present experiment for a few secondary beams are presented in Fig.4 as full lines. None of these distributions is Gaussian-shaped, as would be expected for symmetric fission of a well defined compound nucleus at high excitation energy [100]. The distributions are peaked at about half the nuclear charge of the projectile $Z_{proj}$ and present a noticeable tail. Depending on the projectile and on the impact parameter, the nuclear charge $Z_{prf}$ of the pre-fragment varies over a broad range from $Z_{proj}$ down to around $(Z_{proj} - 20)$ [101]. According to the model calculations described in Section V, in the interaction of $^{219}$Ac at 420 $A$MeV with lead nuclei, events with $Z_{prf} = Z_{proj}$ exhaust about 17% of the fission cross section $\sigma_f$ while events characterized by $Z_{prf} = (Z_{proj} - 10)$ account for 2% of $\sigma_f$. The smaller the nuclear charge $Z_{prf}$ is, the smaller is the nuclear charge $Z_{fiss}$ of the fissioning parent element. The fission-fragment $Z$ distribution is consequently shifted to the left with respect to $Z_{proj}/2$, giving rise to the tail observed in Fig.4. The latter is rather weak which can be explained by the rapidly decreasing cross section for less peripheral collisions and the decreasing fissility of lighter $Z_{fiss}$ elements. In the case of the $^{224}$Th beam, see Fig.4a, in addition to the leftmost tail, shoulders appear at both sides of the central peak (around $Z \approx 36$ and $Z \approx 54$). They correspond to asymmetric fission, originating mainly from electromagnetic-induced processes as well as from the most peripheral nuclear interactions which remove only a few nucleons (mostly neutrons) from the projectile. These reactions induce only low excitation energies in the pre-fragment. No asymmetric component is visible for the lighter beams. This is consistent with the symmetric character of fission at low energies in the pre-actinide region [94, 102, 103]. The individual contributions from electromagnetic- and nuclear-induced interactions, extracted as described in Refs. [94, 97], are shown in Fig.4 by dashed and dotted lines, respectively. The low-energy component peaks at $Z_{proj}/2$ while high-energy fission events are centred at $Z < Z_{proj}/2$.

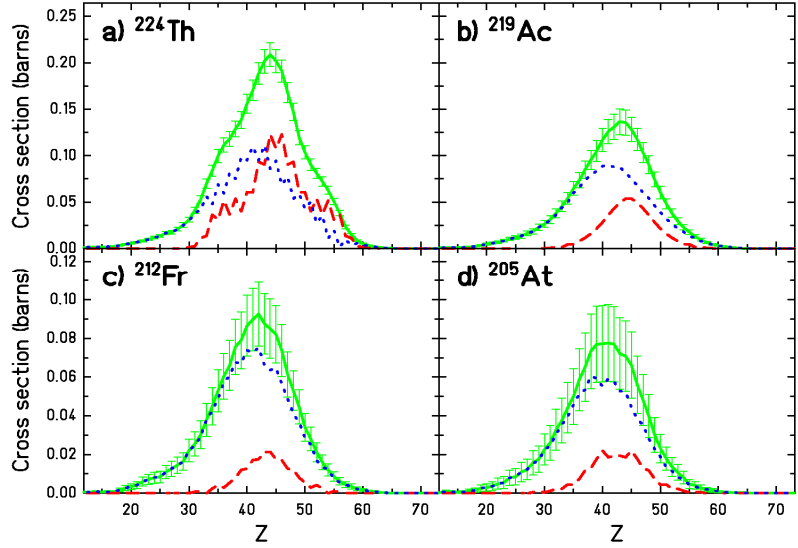

**Figure 4**: (Colour online) Experimental fission-fragment $Z$ production cross sections after interaction of the indicated radioactive projectiles at about 420 $A$MeV with lead nuclei. Full lines encompass all fission events. The experimentally extracted electromagnetic and nuclear contributions are displayed individually by dashed and dotted lines, respectively. For sake of clarity, the error bars of the latter two curves are not included. The error bars include statistical uncertainties.

**b. Classification according to initial excitation energy and fissility**

As discussed above, the pre-fragments formed in the collision between relativistic ions are characterized by small angular momenta [87, 88, 89] but they cover a wide range in $A$, $Z$ and $E^*$ [104, 105]. An accurate investigation of fission induced by a fragmentation reaction therefore requires a classification of the events according to these physical parameters. With the present set-up, a useful sorting can be achieved because the $Z$ of the two fission fragments is measured. Exclusive experiments [104] showed that, in a peripheral collision at relativistic energies, the impact parameter and the nuclear charge of the heaviest fragment measured in an event are strongly correlated. This implies a correlation between the nuclear charge $Z_{prf}$ of the pre-fragment and the impact parameter, and consequently the initial excitation energy. Due to the small probability of charged-particle evaporation by the compound nucleus prior to scission [64, 65, 66], the nuclear charge $Z_{fiss}$ of the fissioning nucleus can be approximated by $Z_{prf}$. Charged-particle evaporation from the fission fragments is scarce as well [64, 65, 66] and $Z_{fiss}$ is equal to $Z_1+Z_2$ in most cases. Hence, the sum $Z_1+Z_2$ can be used as a measure of the initial excitation energy $E^*$ induced in the system (lower $Z_1+Z_2$ values correspond to more central collisions, i.e. higher $E^*$), and of its fissility (lower $Z_1+Z_2$ values imply less fissile pre-fragments). As argued above, the initial shape configuration of the CN pre-fragments studied in this work is nearly spherical because of the choice of the secondary projectiles. Most important parameters are thus under control [34].

α) **The nuclear-charge sum $Z_1+Z_2$**
The significance of a $Z_1+Z_2$ filter is studied in Fig.5. The fission-fragment $Z$ distributions measured for several selections on the sum $Z_1+Z_2$ are superimposed to the integral distribution for the $^{219}$Ac beam. Electromagnetic- and very peripheral nuclear-induced reactions are peaked at $Z_1+Z_2 = Z_{proj}$ [94, 97]. To exclude the influence of structural effects present in low-energy fission from the discussion, we restrict in this work to $Z_1+Z_2$ values smaller or equal to ($Z_{proj} - 2$). For a fixed $Z_1+Z_2$ value, the $Z$ distribution is observed in Fig.5 to be Gaussian-shaped, as expected in fission of a given element at high excitation energies [100]. With decreasing $Z_1+Z_2$, the mean value of the distribution shifts towards smaller values, giving rise to the observed dissymmetry in the integral distribution. This observation corroborates the early interpretation of Fig.4. In addition, it shows based on experimental facts that the sum $Z_1+Z_2$ is a pertinent observable for selecting a range of excitation energy and/or fissility of the CN pre-fragment. Note that events originating from a given secondary beam and a given $Z_1+Z_2$ value correspond to the equivalent of a data set collected in an individual fusion-induced fission experiment. For each of the 45 spherical secondary beams available, 15 up to 20 charge-sum values were exploited in the present work.

β) **The nuclear-charge loss $\Delta Z$**

The initial excitation energy of the CN pre-fragment is to first order given by the difference between $Z_{prf}$ and $Z_{proj}$. That is, for different beams, a given $Z_{prf}$ value, and consequently a given $Z_1+Z_2$ sum, does not correspond to the same range in initial $E^*$. Therefore, the charge loss $\Delta Z = Z_{proj} - (Z_1+Z_2)$, defined as the difference between the nuclear charge of the projectile and the sum of the fission-fragment nuclear charges, is introduced. It permits establishing an excitation-energy scale common to all secondary beams available in the experiment. The nuclear-charge loss is also connected to the fissility of the pre-fragment: A larger $\Delta Z$ implies a less fissile pre-fragment. The latter correlation is projectile-dependent.

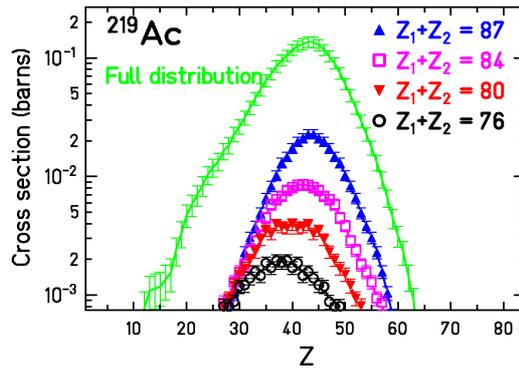

**Figure 5**: (Colour online) Experimental fission-fragment nuclear charge production cross sections measured for a $^{219}$Ac radioactive beam impinging on a lead target. The solid line corresponds to the integral distribution on a logarithmic scale, see Fig.4b), while symbols refer to various gates on $Z_1+Z_2$ as indicated: $Z_1+Z_2 = 87$ (full triangles up), 84 (open squares), 80 (full triangles down) and 76 (open circles).

In summary, the measurements of $Z_1+Z_2$ and $\Delta Z$ provide complementary information. The correlation between $Z_1+Z_2$ and $E^*$ depends on the secondary projectile, while its connection to $Z^2/A$ is absolute. The situation is reversed for $\Delta Z$. For one beam, the range in $E^*$ and $Z^2/A$ selected by either $Z_1+Z_2$ or $\Delta Z$ are correlated. The availability of several beams in the same experiment permits an independent variation of the initial excitation energy and fissility of the compound nucleus. In the remainder of this work, we make use of both the $Z_1+Z_2$ and $\Delta Z$ filters to investigate the potential influence of temperature and fissility on nuclear dissipation. Most previous studies based on fragmentation-induced fission [75, 101, 106, 107, 108] gave, on an event-by-event basis, access only to one of the fission fragments. It was therefore not possible to distinguish the nuclear charges of the various pre-fragments. Sorting the events according to $Z_1+Z_2$ or $\Delta Z$ was feasible in few experiments [68, 109]. However, the stable $^{238}$U beam used in these studies complicates the interpretation as discussed in section I, and the restriction to only one projectile does not permit to vary $E^*$ and $Z^2/A$ independently.

## 2) Experimental fission-fragment nuclear-charge distribution

Since transient effects are expected to be observable at high excitation energies [2, 3, 34, 52] we focus on events with $Z_1+Z_2 \leq (Z_{proj} - 2)$, or equivalently $\Delta Z \geq 2$ [94]. This, as already mentioned, in addition permits to eliminate the influence of nuclear structure on the fission-fragment distribution. The analysis of the low-energy fission component revealed insight into the influence of shell structure and pairing, and is reported in [94, 97, 110, 111].

Figure 6 shows a subset of the measured fission-fragment $Z$ distributions with different selections on $Z_1+Z_2$ (equivalently, $\Delta Z$) for various secondary projectiles. All distributions are Gaussian-shaped; fluctuations are due to limited statistics. Independent of the projectile, the integral yield of the distribution decreases with decreasing $Z_1+Z_2$. That reflects the lower probability for more central collisions as well as the decrease of the fissility of the CN pre-fragment. One also observes that the width of the distribution varies with $Z_1+Z_2$ for a given projectile, as well as with the projectile for a constant $Z_1+Z_2$. These dependences are investigated in the following in detail.

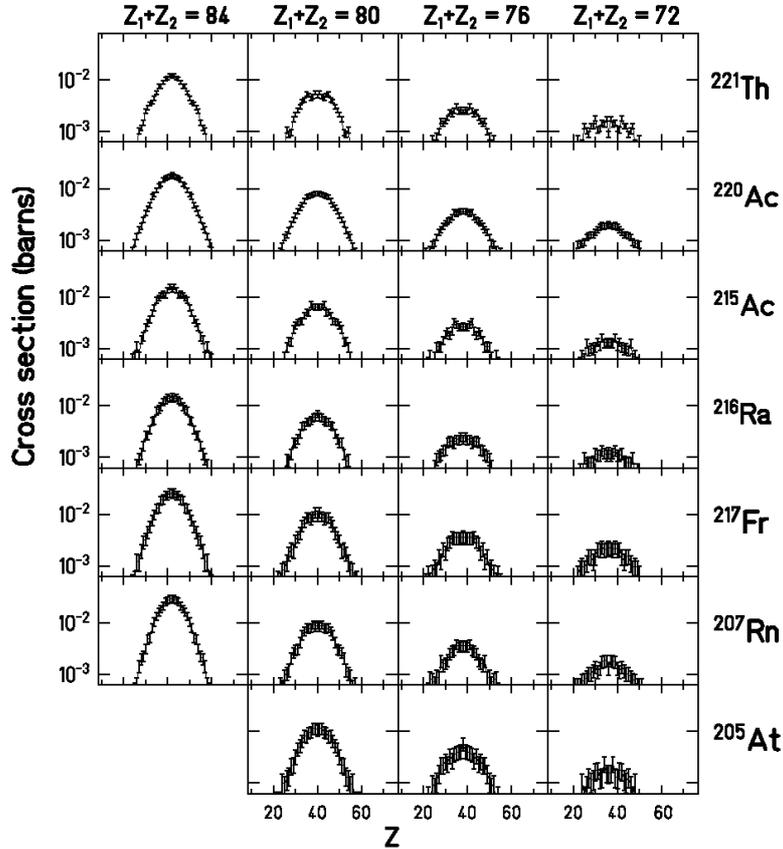

**Figure 6:** Experimental fission-fragment $Z$ distributions for a sample of radioactive beams for different selections on $Z_1+Z_2$. Only events with $Z_1+Z_2 \leq (Z_{proj} - 2)$, equivalently $\Delta Z \geq 2$, are considered.

The width $\sigma_Z$ of the fission-fragment $Z$ distribution was determined for the $Z_1+Z_2$ values with sufficient statistics for each radioactive beam. Its variation as a function of the nuclear-charge sum and nuclear-charge loss is presented in Fig.7 for a sample of secondary projectiles. Over the $Z_1+Z_2$ ($\Delta Z$) interval which is covered in the figure, the initial pre-fragment excitation energy increases nearly linearly [83] from $E^* \approx (100\text{-}130)$ MeV for $Z_1+Z_2 = (Z_{proj} - 2)$ to a saturating value of 550 MeV at $Z_1+Z_2 = 70$ for the most central collisions [105]. Over the same range, the temperature rises from around 2.4 MeV up to 5.5 MeV[3], and the fissility $Z^2/A$ of the compound nucleus decreases from 37.5 to 30.5 (the fissility parameter defined in [112] as $x = Z^2/(A \cdot 50.88 \cdot (1-1.7826 \cdot ((N-Z)/A)^2))$ varies from 0.78 to 0.66).

It is well established that the width $\sigma_A$ of the fission-fragment mass distribution depends on excitation energy, angular momentum and fissility [60, 61, 113, 114, 115, 67]. The influence of these parameters on the nuclear-charge width is expected similar in magnitude, see section I. The increase of $\sigma_Z$ with decreasing (increasing) $Z_1+Z_2$ ($\Delta Z$) observed in of Fig.7 is, for largest part, caused by the increase in $E^*$ as more and more nucleons are abraded from the projectile. The manifestation of the influence of the fissility is less obvious. According to the parameterization by Rusanov et al. [60] the variation of the stiffness $d^2V/d\eta^2$ with fissility reaches a maximum around $Z^2/A \approx 34$, and $d^2V/d\eta^2$ depends only weakly on $Z^2/A$ for $32 \leq Z^2/A \leq 34$ (see e.g. Fig.8 of [60]). The model calculations described in section V suggest that, for the pre-fragments studied in the present work, the fissility distribution is centered around $Z^2/A \approx 34$ with a tail down to $Z^2/A \approx 30$-31. Based on (Eq.2), we therefore expect a slight increase of $\sigma_Z$ with increasing $Z^2/A$, or equivalently, a decrease of $\sigma_Z$ with decreasing (increasing) $Z_1+Z_2$ ($\Delta Z$). Tracing the potential influence of fissility is nonetheless difficult. Not only is its influence expected to be weak, but the experimental information is limited to $Z_{proj}$, $N_{proj}$ and $Z_1+Z_2$. The fissility of the fissioning nucleus which can be estimated from the measured $Z_1+Z_2$ might not be sufficiently accurate to determine the fissility of the fissioning nucleus due to intermediate particle (mostly neutron) evaporation. In addition, the width of the excitation energy window selected by the available filters might wash out fissility effects. For instance, for a given value of $\Delta Z$, $\sigma_Z$ is found to be nearly independent of the fissility of the projectile. The present experiment, therefore, does allow only for a general conclusion of the dependence of $\sigma_Z$ on fissility. Namely, the data suggest that there is no strong influence of $Z^2/A$ on the width of the fission-fragment $Z$ distribution, and support a dominant influence of the excitation energy.

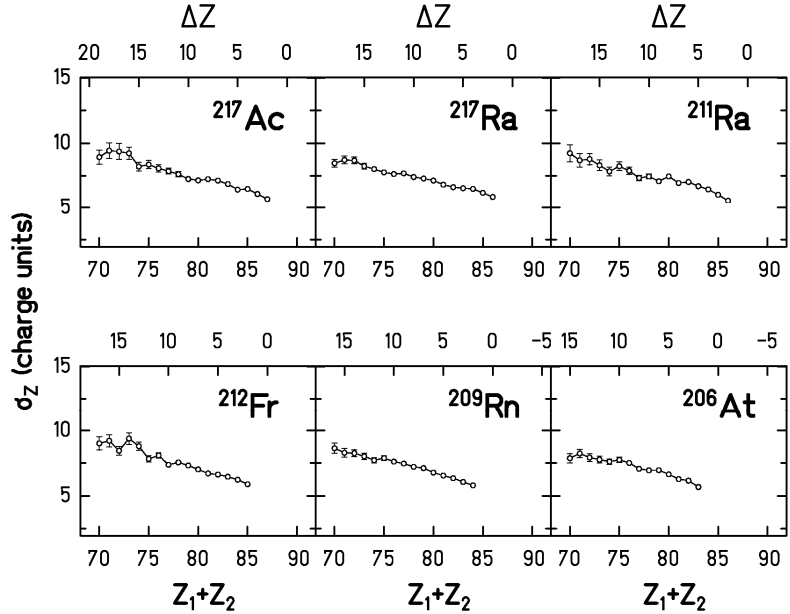

---

[3] The temperature given here corresponds to an average value over the various isotopes contributing to a given $Z_1+Z_2$ for typical secondary beams. Excitation energy and temperature were related to each other according to $E^* = a \cdot T^2$, where the level density parameter $a$ is calculated following Ignatyuk's prescription [A.V.Ignatyuk et al., Yad. Fiz. 21 (1975) 1185].

**Figure 7:** Width $\sigma_Z$ of the fission-fragment Z distribution as a function of $Z_1+Z_2$ (lower axis) and $\Delta Z$ (upper axis) for a selection of radioactive beams. Only those events with $Z_1+Z_2 \leq (Z_{proj} - 2)$, equivalently $\Delta Z \geq 2$, are considered.

The most commonly used observables in the investigation of nuclear fission are cross sections, fission-fragment masses, kinetic energies and angular distributions as well as light-particle and γ-ray multiplicities. There exits little information on nuclear-charge distributions, mainly due to the difficulty of measuring Z with good resolution. To compare the results of the present work with previous measurements, a simple procedure was therefore used to estimate $\sigma_Z$ from experimental data on $\sigma_A$. This procedure and the detailed comparison of the present and previous results are reported in Appendix I. The measurements of this work are found to agree with those derived in the past *via* various techniques.

### V. Manifestation and magnitude of transient effects

**1) Model calculations**

To extract information on fission transient effects from experiment requires the support of model calculations. Peripheral collisions between heavy ions at relativistic energies proceed *via* well-separated stages and can be described within an abrasion-ablation scenario [79, 80]. In the present work, we use the ABRABLA reaction code [80, 83, 116, 117] which is based on the participant-spectator picture.

A peripheral nucleus-nucleus collision at energies well above the Fermi velocity can be described as a cascade of nucleon-nucleon collisions. This cascade is located almost exclusively in the overlap volume of the projectile and the target. The nucleons in this zone are the so-called participants of the reaction. The nucleons outside this volume, the so-called spectators, are mostly unaffected by the cascade and proceed moving almost undisturbed. In a simple geometric picture, the resulting projectile-spectator (or pre-fragment) can be seen as the initial projectile that lacks the part sheared off by the target (cf. Fig.1). Its mass depends on the impact parameter. In the ABRABLA code, for a given mass loss, protons and neutrons are randomly removed from the projectile, and the N/Z ratio of the pre-fragment is subject to statistical fluctuations given by the hyper-geometrical distribution [118]. The excitation energy $E^*$ induced in the pre-fragment is partly caused by the single-particle energies of the holes created by the abraded nucleons. Experimental data [83] show that, on average, an excitation energy of 27 MeV per abraded nucleon is induced into the recoiling spectator. The angular momentum of the pre-fragment is obtained from the sum of the angular momenta of the removed nucleons on the basis of the shell model [87]; its root-mean-squared value varies from 10 to 20 ℏ.

A second stage in the ABRABLA code accounts for multi-fragmentation-like phenomena which occur due to thermal instabilities if the temperature of the pre-fragment exceeds ≈ 5.5 MeV [119, 105]. This so-called break-up stage is expected to be very fast and results in the simultaneous emission of nucleons and clusters. The N/Z ratio of the initial pre-fragment[4] is conserved, and the amount of released energy is parameterised on the basis of experimental information [105]. The omission of multi-fragmentation-like phenomena can lead to an overestimation of fission at very high temperature, and result in an artificially large dissipation strength [71].

---
[4] Throughout this paper, we identify the pre-fragment with the excited product resulting from either abrasion or abrasion followed by a multi-fragmentation-like stage, depending on whether the latter occurs.

Once the temperature decreases below 5.5 MeV, the sequential de-excitation phase, or ablation, of the remaining excited pre-fragment is treated within the statistical model. The competition between evaporation of light particles (*n*, *p*, *d*, *t*, $^3$He, α) and fission is governed by the available phase space for each channel. Particle emission probabilities are calculated according to the Weisskopf-Ewing formalism [120]. Special care is taken to describe the fission channel [133]. The fission probability is calculated at each evaporation step as a function of mass, charge, excitation energy *and* time using the analytical expression of the fission decay-width $\Gamma_f(t)$ derived by Jurado *et al.* [121, 122] for initially spherical compound nuclei. This expression was shown to account for the early inhibition of fission caused by relaxation effects in a realistic way. The result was found to be very close to that obtained by the numerical solution of the Fokker-Planck equation. In this respect, the treatment of the competition between fission and evaporation in ABRABLA includes in a highly realistic way [122] the dynamical effects originally evidenced by the solution of the equation of motion. The analytical formula derived for $\Gamma_f(t)$ in Refs. [121, 122] avoids the deficiencies inherent to previous prescriptions (see discussion in Ref. [122]). Aside from the description of dissipation effects, the critical ingredients of any de-excitation code are the fission barrier $B_f$ and the ratio of the level-density parameter at the saddle point to that at the ground state $a_f/a_n$. The macroscopic part of the fission barrier is taken from Sierk's finite-range liquid-drop model [123] and the level-density parameter is calculated including volume and surface dependencies as proposed by Ignatyuk *et al.* [124]. Theory [125] as well as experiment [126] show that these parameters are highly realistic. In case the system undergoes fission along the de-excitation cascade, the mass and nuclear charge distributions of the fragments are computed from the semi-empirical parameterisation developed by Benlliure *et al.* [127]. Recently, McCalla and Lestone [51, 52] emphasized the critical influence of several ingredients of the statistical model. They recommend the use of a modified expression for the Kramers fission decay-width that takes into account the stationary collective states of the system about the ground-state and the saddle point (referred as to the Strutinsky factor [128]), the temperature dependence of the location of the saddle point and the orientation degree of freedom. These features are not included in the version of ABRABLA used for the present study. The systems measured in our experiment have low angular momenta, and orientation effects are negligible. The modification of the saddle point location with temperature noticeably affects the fission cross sections and pre-scission particle multiplicities [52]. We have checked that the influence on the *Z*-width is weak. The necessity of including the Strutinsky factor is less obvious. It was recently shown [129] that the application of the Strutinsky factor acts in a very similar way as a reduction of the effective value of the $a_f/a_n$ ratio. In [126] cumulative fission probabilities were measured for neighbouring nuclei. This is the most direct and reliable way to extract information on the $a_f/a_n$ ratio. In the model they used, Jing *et al.* [126] did not explicitly introduce the Strutinsky factor, i.e. the value they extracted for $a_f/a_n$ necessarily includes this factor. The $a_f/a_n$ ratio obtained in [126] was found to be in close agreement with the prescription by Ignatyuk *et al.* [124]. Independent of any theoretical justification, the use of the parameterisation of Ignatyuk *et al.* in the present work is thus equivalent to the use of the empirical effective $a_f/a_n$ ratio calibrated by the data of [126]. It permits to account for the Strutinsky factor in an effective way. Additionally, it should be emphasized that the *Z*-width has been shown to be a direct and robust signature of transient effects [69]; its dependence on uncertain model parameters is weak. The ABRABLA code is able to reproduce data of different kinds and measured in various reactions (see e.g. [75, 76, 97, 85, 127, 130, 117]).

As discussed in the previous section, experimental observations suggest that $Z_{prf} \approx Z_{fiss} \approx Z_1+Z_2$. Figure 8 shows that the ABRABLA model calculations support this ansatz. The calculated correlation between $Z_1+Z_2$ and $Z_{fiss}$ (left panel) excludes post-scission light-charged particle evaporation in most cases. The dependence of $Z_{fiss}$ on $Z_{prf}$ is strong as well (central panel), indicating the emission of few charged particles prior to scission only for the smallest $Z_{prf}$ i.e. the highest excitation energies. The sum $Z_1+Z_2$ is therefore as expected correlated with the initial excitation energy $E^*$ of the CN pre-fragment (right panel). The saturation of $E^*$ at $\approx$ 550 MeV originates from the onset of multi-fragmentation-like processes above temperatures around 5.5 MeV [105]. The significance of the nuclear-charge loss $\Delta Z$ filter derived from the model calculations is presented in Fig.9. For $\Delta Z = 5$ the initial excitation energy is centred around 200 MeV, while $\Delta Z = 10$ corresponds to $E^* \approx 350$ MeV, almost independent of the secondary projectile. Again, the saturation at $E^* \approx 550$ MeV is caused by the onset of multi-fragmentation.

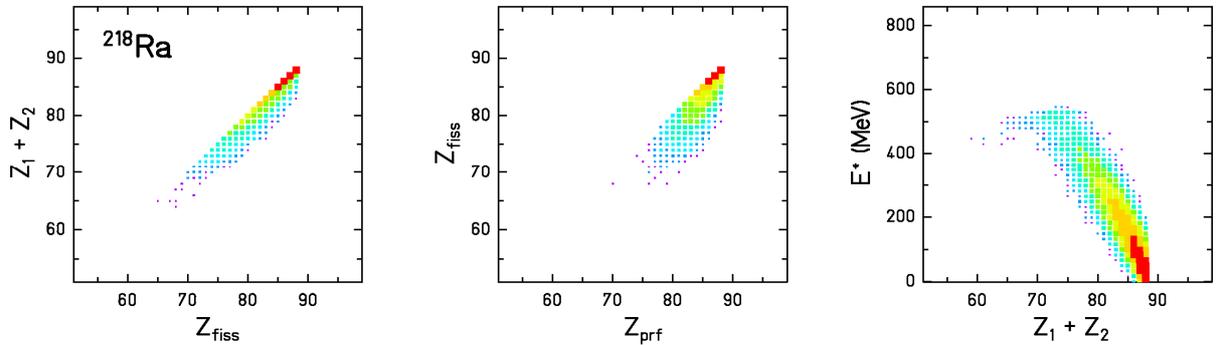

**Figure 8**: (Colour online) Calculated correlation between the charge $Z_{fiss}$ of the fissioning nucleus and the sum $Z_1+Z_2$ of the two fission-fragment nuclear charges (left), the charge $Z_{prf}$ of the pre-fragment and $Z_{fiss}$ (middle), as well as $Z_1+Z_2$ and the initial excitation energy $E^*$ of the pre-fragment (right). The calculations were done for a $^{218}$Ra beam using a dissipation strength $\beta = 5 \cdot 10^{21}$s$^{-1}$. Only nuclear-induced fission events are considered. The displayed colour scales logarithmically with the observed intensity.

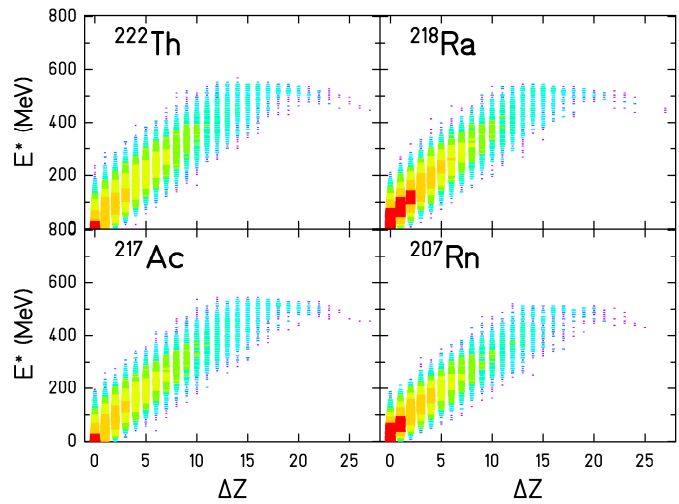

**Figure 9**: (Colour online) Calculated correlation between the nuclear-charge loss $\Delta Z$ and the initial excitation energy $E^*$ for typical secondary beams. The calculations were done with a dissipation strength of $\beta = 5 \cdot 10^{21}$s$^{-1}$. Only nuclear-induced fission events are considered. The displayed colour scales logarithmically with the observed intensity.

Although the correlation between $Z_1+Z_2$, equivalently $\Delta Z$, and the initial excitation energy is well established, the accuracy of the measurement of $E^*$ by a selection on either $Z_1+Z_2$ or $\Delta Z$ is limited. The model calculations suggest that, depending on the system and gating, the selection of $E^*$ fluctuates by (15-40)% around its mean value. This uncertainty is mainly due to the lack of information on the fission-fragment mass in the present work. Similar attempts for classifying the events with respect to $E^*$ were made in previous works performed in normal kinematics by exploiting either the mass loss [131, 132] or the total light-particle multiplicity [78]. At best, an accuracy of (10-30)% in $E^*$ was reached.

**2) Evidence for the manifestation of transient effects at high excitation energy**

As argued previously, the width of the fission-fragment $Z$ distribution is expected to be primarily governed by the dynamics in the quasi-bound region. Both the fission delay ($\tau_{trans}$) and the magnitude of the fission probability (related to $\Gamma^K$) matter *a priori*. A calculation omitting dissipation effects completely by using the Bohr-Wheeler fission decay-width has indeed shown to be inappropriate for describing the present measurements, see Fig. 4 of [34]. To isolate the influence of transient effects, the data are confronted with two types of calculations: A calculation based on Kramers' *time-independent* fission decay-width $\Gamma^K$ [11] and a so-called transient-type calculation using the realistic *time-dependent* formula for $\Gamma_f(t)$ derived in [121, 122]. Both options are implemented in ABRABLA [133]. The difference, if any, between such two calculations is to be ascribed to the transient delay. For a sample of spherical projectiles, the behaviour of the experimental $\sigma_Z$ as a function of $Z_1+Z_2$ is compared in Fig.10 to Kramers- and transient-type predictions. In both calculations, the dissipation strength was fixed at $\beta = 4.5 \cdot 10^{21} \mathrm{s}^{-1}$. It can be seen that the slope of $\sigma_Z$ with decreasing nuclear-charge sum depends strongly on the inclusion of transient effects in the calculation. This observation is valid independent of the value used for $\beta$. Namely, it is *not* possible to reproduce the trend in $\sigma_Z$ as a function of $Z_1+Z_2$ with the time-independent $\Gamma^K$ and an artificial increase of $\beta$ with decreasing $Z_1+Z_2$. This is illustrated in Fig.10 for the $^{223}$Ac projectile with several calculations which use the decay-width $\Gamma^K$ and values of $\beta$ between $7 \cdot 10^{21} \mathrm{s}^{-1}$ and $20 \cdot 10^{21} \mathrm{s}^{-1}$. The present data set can therefore not be explained by a calculation which omits transient effects but assumes a temperature-dependent $\beta$. Note that an artificial increase of the dissipation strength also modifies the total fission cross section up to a point at which it is inconsistent with the measured value [133]. The necessity of including transient effects in the description of the experimental slope of $\sigma_Z$ as a function of $Z_1+Z_2$ indicates that the influence of $\tau_{trans}$ on $\sigma_Z$ dominates over that of the magnitude of the quasi-stationary fission decay-width. This can be understood from the fact that, during a large part of the transient delay, fission is completely inhibited [122]. Time is available for evaporation, which reduces the temperature at the saddle point. This temperature determines the $Z$-width. Meanwhile, the Kramers' factor leads to an overall reduction of the fission decay-width. It is worth noting that the Kramers-type calculations steadily diverge from the data as $Z_1+Z_2$ decreases, indicating that transient effects manifest themselves more strongly at high excitation energies. This progressive departure from the data is due to the decrease of the saddle-point temperature $T^{sad}$ due to transient effects which becomes particularly strong at high excitation energies. At low temperatures, particle emission times are comparable to the transient time, which hampers evaporation during $\tau_{trans}$. With increasing $E^*$, the time for emitting a light particle decreases much faster than $\tau_{trans}$ does [9], and the number of particles which are evaporated during the fission delay increases.

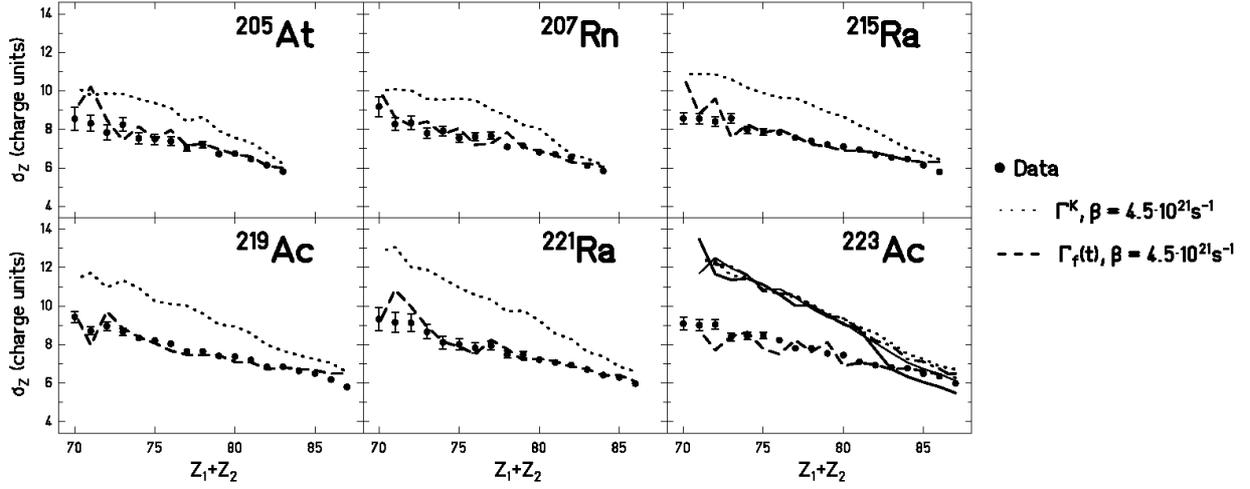

**Figure 10:** Width $\sigma_Z$ plotted as a function of $Z_1+Z_2$ for a set of radioactive beams. The data (black dots) are compared to Kramers- (dotted) and transient- (dashed) type calculations, both assuming $\beta = 4.5 \cdot 10^{21} s^{-1}$. For the $^{223}$Ac secondary projectiles, additional Kramers-type calculations are shown with $\beta = 7 \cdot 10^{21} s^{-1}$ (dash-dotted), $\beta = 10 \cdot 10^{21} s^{-1}$ (full thin) and $\beta = 20 \cdot 10^{21} s^{-1}$ (full thick). The staggering that can be observed in the calculations is due to statistical fluctuations only. Experimental error bars smaller than the symbols are not shown.

The inhibition of fission at high temperatures is further illustrated in Fig.11 where experimental and calculated values of $\sigma_Z$ are shown for a wide range of systems gated on either $Z_1+Z_2$ or $\Delta Z$. The Z-width is displayed as function of the neutron number $N_{proj}$ of the projectile for all Ac and Ra secondary beams available in the experiment. We will use this type of presentation in the following, because it allows a compact view of the large variety of systems. Similar to Fig.10 the predictions of Kramers-type calculation and those obtained with the $\Gamma_f(t)$ approximation of [121, 122] are close to each other at large $Z_1+Z_2$ (low $\Delta Z$), and progressively differ from each other with decreasing nuclear-charge sum (increasing nuclear-charge loss). The discrepancy between the calculations neglecting transient effects and the data above $\Delta Z \approx 2$ suggests that the $\sigma_Z$ signature becomes a very particularly relevant signature of transients at $E^* \approx 130$ MeV, or equivalently initial temperatures above about 2.4 MeV.

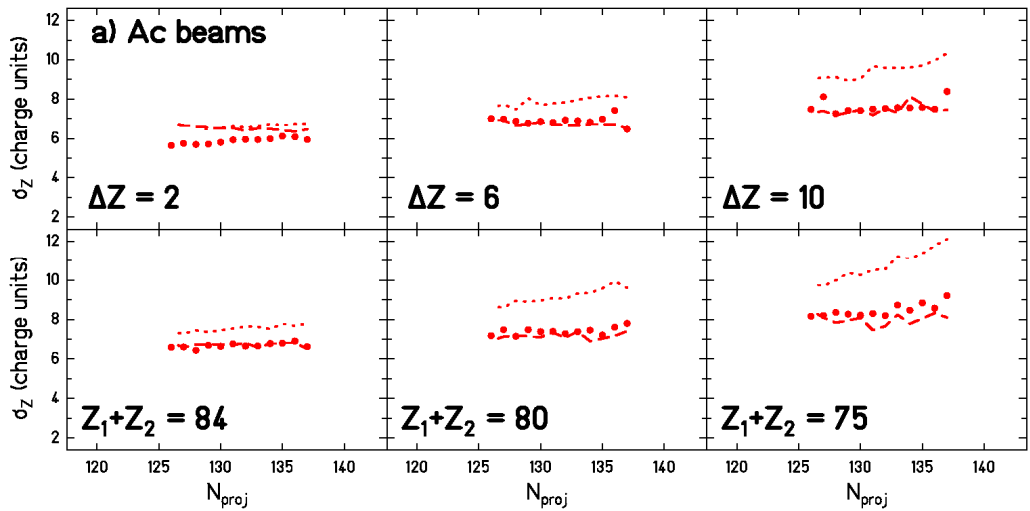

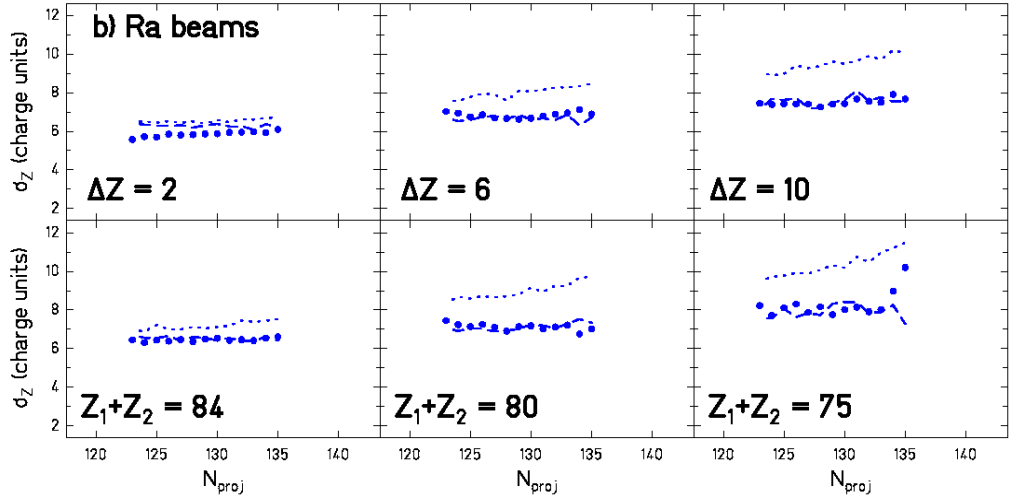

**Figure 11:** Width $\sigma_Z$ plotted as a function of $N_{proj}$ for the Ac (top) and Ra (bottom) radioactive projectiles gated on $\Delta Z$ or $Z_1+Z_2$ as indicated. The data (dots) are compared to Kramers- (dotted) and transient- (dashed) type calculations, both assuming $\beta = 4.5 \cdot 10^{21} s^{-1}$. Staggering of the calculations is due to statistical fluctuations only. Experimental error bars smaller than the symbols are not shown.

In conclusion, relaxation effects prior to the saddle point are observable and are important for the description of fission. This result is at variance with some previous statements [23]. Recent macroscopic [24] and microscopic [9] calculations support this point of view. The unambiguous manifestation of transients requires well suited initial conditions (namely, a compact shape configuration and sufficiently large excitation energies), the availability of a specifically sensitive experimental observable and realistic model calculations.

### 3) Dissipation strength and transient delay

#### a. Nuclear dissipation coefficient

To determine the magnitude of the dissipation strength, the experimental $Z$-widths were compared to calculations performed with the analytical $\Gamma_f(t)$ approximation of Refs. [121, 122] for different values of the dissipation strength from the under-damped ($\beta = 1 \cdot 10^{21} s^{-1}$) to the over-damped ($\beta = 7 \cdot 10^{21} s^{-1}$) regime. As an example, Fig.12 shows the results obtained for the Ac and Ra secondary projectiles for a few selections on $Z_1+Z_2$ and $\Delta Z$. A similar comparison of the calculations with all the data measured in the present work shows that $\beta = 4.5 \cdot 10^{21} s^{-1}$ is best suited, independent of the beam and the nuclear-charge sum or loss gate. To assess the reliability of the extracted value for $\beta$, we investigated the influence of most critical model parameters on its value. Varying the fission barrier $B_f$, the $a_f/a_n$ ratio and the temperature at which multi-fragmentation sets in by 10% changes $\sigma_Z$ by 12% at worst. That leads us to conclude that the present measurement is best described assuming an over-damped motion with $\beta = (4.5 \pm 0.5) \cdot 10^{21} s^{-1}$. The fact that the initial pre-fragment is nearly spherical and the specific sensitivity of $\sigma_Z$ to the pre-saddle region implies that the value extracted for $\beta$ is an estimate of the dissipation strength at small deformation.

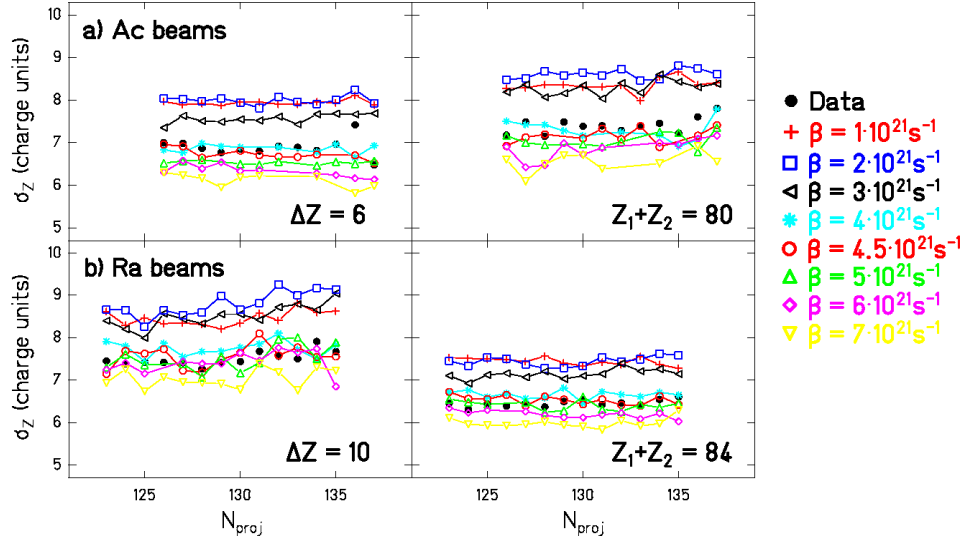

**Figure 12:** (Colour online) Width $\sigma_Z$ plotted as a function of $N_{proj}$ for the Ac (top) and Ra (bottom) radioactive projectiles gated by $\Delta Z$ or $Z_1+Z_2$. The data (dots) are compared with transient-type calculations assuming different values for the dissipation strength $\beta$ as indicated by the symbols joined by lines.

### b. Transient time

According to the dynamical picture of fission, the time the system needs to establish quasi-equilibrium above the saddle point is closely related to the magnitude of the dissipation strength in the corresponding deformation regime. The analytical approximation of the fission-decay width $\Gamma_f(t)$ derived in [121, 122] is based on the numerical solution of the Fokker-Planck equation [2, 3] for an initially spherical nucleus. For such an initial configuration, the relation between $\tau_{trans}$ and $\beta$ was parameterised by Grangé and co-workers [2, 3]. For the over-damped regime it reads:

$$\tau_{trans} = \beta/2\omega_g^2 \cdot ln(10B_f/T) \qquad (Eq.3)$$

where $\omega_g$ is the oscillator frequency at the ground-state and $T$ is the initial temperature of the decaying system. Since the initial shape configuration in the present work matches the one of Grangé et al. [2, 3] and Jurado et al. [121, 122], the value of the transient time involved along the de-excitation chain can be extracted from the calculations performed with $\beta = 4.5 \cdot 10^{21} s^{-1}$ in section V.1. The so-obtained transient time is presented as a function of $\Delta Z$ and $Z_{proj}^2/A_{proj}$ in Fig.13 for a sample of secondary projectiles. Within the error bars, $\tau_{trans}$ is approximately constant around an average value of $(3.3\pm0.7) \cdot 10^{-21}$s. This estimate is in very good agreement with recent microscopic calculations in the corresponding temperature range (see Fig.3 in [9] for 3 MeV $< T <$ 5 MeV).

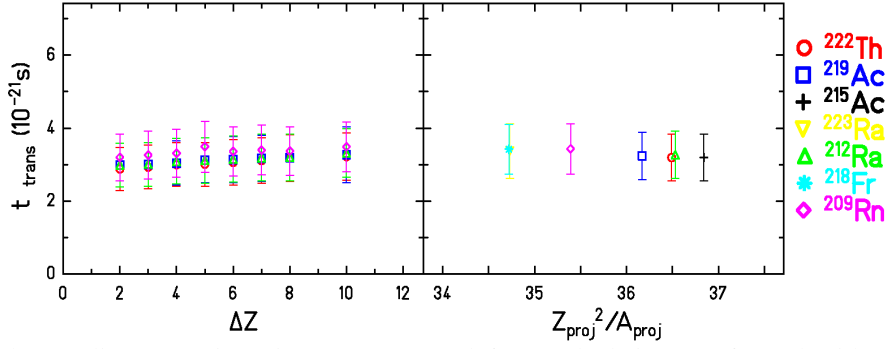

**Figure 13:** (Colour online) Transient time $\tau_{trans}$ extracted from calculations performed with $\beta = 4.5 \cdot 10^{21} s^{-1}$, displayed as a function of $\Delta Z$ (left) and $Z_{proj}^2/A_{proj}$ (right). Only a sample of radioactive projectiles is shown for sake of clarity.

### c. Temperature and fissility dependence

The $Z$ distributions measured after the fragmentation-induced fission of the 45 spherical projectiles available in the present work can be described by a unique ($\beta$, $\tau_{trans}$) combination, independent of the selection on the nuclear-charge sum or loss. That is, we find no evidence for a temperature or fissility dependence of pre-saddle dissipation strength.

A temperature- and fissility-independent $\beta$ does not imply a $T$- and $Z^2/A$-independent $\tau_{trans}$. As Fig.13 demonstrates, an increase of the transient time for decreasing $T$ or $Z^2/A$ (via the increase of $B_f$ in the latter case) is not observed. This is partly explained by the expected logarithmic dependence of $\tau_{trans}$ on $T$ and $B_f$ which is weaker than its linear increase with $\beta$ in the over-damped regime, see (Eq.3). Furthermore, the variation of the transient time with temperature as predicted by macro- [3] and microscopic [9] calculations for the conditions of the present work i.e. $\beta = 4.5 \cdot 10^{21} s^{-1}$ and 3 MeV $< T <$ 5 MeV, is limited to $\approx$ (0.5-1)$\cdot 10^{-21}$s around its mean value. This is within the experimental error bars of our work which is therefore not sensitive to it. There exist very few data probing the influence of the fissility on $\tau_{trans}$. The question on an observable effect is still debated [46, 115].

The above statement on the absence of a $T$- and $Z^2/A$-dependence of $\beta$ and $\tau_{trans}$ is valid within the limited accuracy of the selection of $E^*$ and $Z^2/A$ in the present experiment. Nonetheless, the study excludes a strong influence of temperature and fissility on nuclear dissipation phenomena at small deformation.

Summarizing, the ability of the calculations using a realistic expression of $\Gamma_f(t)$ [121, 122] and a dissipation strength $\beta = 4.5 \cdot 10^{21} s^{-1}$ to reproduce the measured $Z$-widths, independent of the initial excitation energy and fissility, is illustrated in Fig.14 for a large sample of the data. The discrepancy observed for some Ac and Th projectile isotopes for $\Delta Z = 2$ is due to the persistence of low-energy fission - leading to either narrow symmetric or wide asymmetric nuclear-charge splits [94, 97]. We emphasize again that the availability of various beams combined to the measure of $Z_1+Z_2$ (equivalently $\Delta Z$) allows varying the initial excitation energy and fissility of the pre-fragment independently. For instance, $\Delta Z = 6$ corresponds to $E^* \approx$ 250 MeV independent of the projectile, whereas the nuclear composition of the CN pre-fragment strongly depends on it: e.g. $Z_{prf} = 83$ and $Z_{prf} = 79$ for Ac and At beams, respectively. A given nuclear-charge sum, e.g. $Z_1+Z_2 = 84$, refers to a similar pre-fragment fissility, while $E^*$ can noticeably differ depending on the projectile: around 125 MeV for a $^{208}$Rn beam and nearly twice this value for $^{222}$Th. It is remarkable that the model is able to describe well the experiment for such a large range of elements, isotopes, temperatures and fissilities.

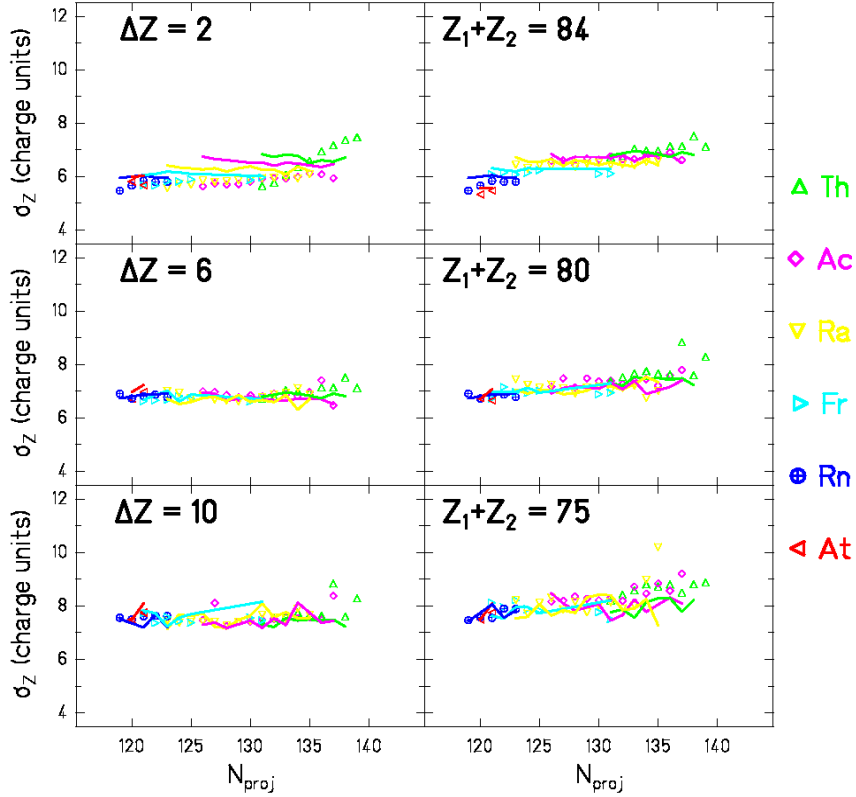

**Figure 14:** (Colour online) Width $\sigma_Z$ plotted as function of $N_{proj}$ gated on different values of $\Delta Z$ or $Z_1+Z_2$. All available secondary projectiles are included. The data (symbols) are compared to transient-type calculations with $\beta=4.5\cdot10^{21}$s$^{-1}$ (full lines).

## VI. Discussion

In the following, the results of this work are discussed in comparison with a selection of representative experimental and theoretical studies which can be found in literature. Before doing so, some important aspects inherent to the interplay between the choice of physical parameters in the calculations and model-dependence of the extracted results are reviewed. This discussion is included in an attempt to understand previous and present findings.

**1) Physics interplay and model dependence**

The nature and magnitude of nuclear dissipation and its dependence on physical parameters has been a long-standing point of controversy in literature. Fröbrich et al. [47] suggested a small dissipation coefficient $\beta^{pre-sad}$ in the pre-saddle region and an increasing strength $\beta^{post-sad}$ along the saddle-to-scission descent as an explanation for observed large neutron pre-scission multiplicities without a dramatic decrease of the corresponding fission cross section. Hofman et al. [134] introduced a dissipation strength $\beta$ which increases rapidly with temperature and was thus able to describe simultaneously GDR γ-ray spectra and evaporation residue cross sections. The inconsistency between these two approaches was pointed out in Ref. [135] where experimental data are described equally well with either a deformation- or a temperature-dependent dissipation strength. The influence of the nuclear composition on the manifestation of dissipation in fission was discussed in Ref. [136]. A deformation dependence of $\beta$ might be masked by the decreasing fissility of light compound nuclei which are characterized by a shorter descent between the saddle and the scission point. The later results in shorter saddle-to-scission times $\tau_{ss}$. The contribution from $\beta$ values characteristic for large deformations is thus reduced. The critical influence of the fissility on the sensitivity to nuclear dissipation and how this is related to the chosen observable has also been investigated recently [91, 92, 93]. Microscopic calculations [18] predict both, a deformation- and a temperature-dependence of dissipation. However, these dependences are weaker than what was extracted in Refs. [47, 134] because of interference between the two. The disagreement among existing theories, see e.g. [93], further contributes to the problems in the interpretation of the experimental data.

The uncertainties of the parameters entering into the statistical model analysis have a large impact on the conclusion on both the magnitude of nuclear dissipation and the very existence of transient effects. The sensitivity of $\sigma_{ER}$ and $\sigma_f$ on the $a_f/a_n$ ratio [137, 138, 75, 50, 51] and on the fission barrier $B_f$ [139, 140, 141] is particularly important. As demonstrated in Refs. [69, 76] an unrealistic $a_f/a_n$ ratio in a calculation which neglects transient effects can result in cross sections similar to those derived from a calculation with more elaborate estimates of $a_f/a_n$ and $\tau_{trans} \neq 0$ (see also below). As soon as light-charged particles (LCPs) or GDR $\gamma$-rays are considered, the limited knowledge of the corresponding partial decay widths [142, 143, 144, 64] complicates further the interpretation of the data. The inadequate description of several additional aspects of fission process was discussed recently in Refs. [52, 145].

## 2) Comparison of conclusions on dissipation and time scales in fission

### a. Fusion-induced fission

Fusion-induced fission is the reaction mechanism most often used for probing nuclear dissipation. An exhaustive list of references would be difficult to establish, thus we limit ourselves in the following to representative works. The compound nucleus formed in a fusion reaction is well-defined in nuclear mass and nuclear charge. The range in excitation energy is limited and given by the reaction $Q$-value and the bombarding energy [70] while the angular momentum can span a broad range. From a compilation of data on pre-scission neutrons, LCPs and $\gamma$-rays measured in fusion-induced fission reactions, Thoennessen *et al.* [146] concluded that there exists a "threshold excitation energy for the appearance of non-statistical fission". They located this threshold around $E^* = (50-80)$ MeV for systems similar to those studied in that work. A slightly lower limit around $E^* = 40$ MeV was suggested by Siwek-Wilcynska *et al.* [72] based on cross section data. In Ref. [147] Moretto *et al.* concluded that there is no room for pre-saddle transients within the range 60 MeV $< E^* <$ 140 MeV. This statement was debated [148] and it was recognized later [149] that pre-saddle delays smaller than $10 \cdot 10^{-21}$s can not be ruled out in the quoted excitation-energy interval. Most compound nuclei produced in fusion reactions have excitation energies below 120 MeV, which is certainly near the threshold at which transient effects themselves manifest strongly; at lower excitation energies the particle emission times are comparable to the delay of fission. This, together with potential entrance-channel effects and the differences in the available observables and modeling, might partly explain why a number of excitation functions up to moderate $E^*$ can be described without taking transient effects into account (see e.g. [147, 150, 151]), while the evidence for a finite fission delay down to low temperatures is found too, see e.g. [152, 153]. Some of these aspects are discussed in detail in the following with illustrative examples.

Based on the detection of the evaporated neutrons, a systematic study [43, 154] scanning a wide range of excitation energies and fissilities suggested a minimum for the so-called dynamical time $\tau_{trans}+\tau_{ss}$ at $(30-35)\cdot10^{-21}$s for $E^*$ up to 150 MeV. Approaches based on the detection of LCPs [65] and γ-rays [155] resulted in an upper limit of $(10-20)\cdot10^{-21}$s for $\tau_{trans}$ and $(20-80)\cdot10^{-21}$s for $\tau_{ss}$. Depending on the initial temperature, either the pre- or the post-saddle stage was found to dominate. Over a fissility range similar to that of the present work, but at a lower excitation energy $E^* \approx 60$ MeV, Saxena et al. [46] extracted a nearly constant transient time of around $10\cdot10^{-21}$s. In Ref. [156] the value of $\tau_{trans}$, deduced for the $^{179}$Re compound system, was found to increase from $15\cdot10^{-21}$s up to $45\cdot10^{-21}$s if $E^*$ decreased from 254 MeV to 112 MeV. Tentatively, one may suggest a slightly shorter fission delay for the systems investigated in the present work because of the temperature of the CN pre-fragment which is higher on average. According to microscopic calculations [9], for $\beta \approx 5\cdot10^{21}$s$^{-1}$, the variation of $\tau_{trans}$ with $T$ becomes strong below 3 MeV. Such a dependence of $\tau_{trans}$ on $T$ implies larger values for the mean extracted transient time in previous works (with $T$ well below 3 MeV) compared to the present one (with $T$ above 3 MeV on average). Nonetheless, the variety of $a_f/a_n$ ratios used in literature might also be a source of diverging conclusions, see e.g. [46, 157, 158]. We note also that in some studies [155, 156] the overall reduction of $\Gamma_f(t)$ caused by Kramers' $K$ factor is not included. This omission mainly affects the evolution of compound nuclei with the lowest excitation energies, because it enhances the asymptotic value of $\Gamma_f(t)$. Events with large $E^*$ have a very high fission probability anyway. The contribution from low $E^*$ events is artificially larger, so that the effective excitation energy is somewhat lower.

Using a combination of a dynamical and a statistical code, Fröbrich et al. [47] obtained, as mentioned above, a deformation-dependent dissipation with a constant strength inside the saddle of $\beta^{pre-sad} = 2\cdot10^{21}$s$^{-1}$ and an increasing strength $\beta^{post-sad}$ along the descent to scission up to $30\cdot10^{21}$s$^{-1}$. Their analysis relied on pre-scission particle multiplicities, and a mixture of pre- and post-saddle contributions cannot be ruled out. That is, a value of $\beta^{post-sad}$ which is too large can compensate for a value of $\beta^{pre-sad}$ which is too small. The authors suggested that the shape-dependence of $\beta$ is probably more gentle. Advanced dynamical calculations [159, 160] promote a large dissipation $\beta^{pre-sad} = (3-7)\cdot10^{21}$s$^{-1}$ at small deformation. Experimentally, the sensitivity of the GDR γ-ray spectrum to the deformation of the emitting source [161] was exploited to disentangle pre- and post-saddle effects. For the small deformation regime, Dioszegi et al. obtained $\beta^{pre-sad} \approx 6\cdot10^{21}$s$^{-1}$ [162]. The larger value of up to $\beta \approx (20\pm6)\cdot10^{21}$s$^{-1}$ extracted in Refs. [163, 164] from similar GDR data is averaged over the entire fission path, and may therefore be understood as a signature of a strong $\beta^{post-sad}$ component. To be more sensitive to the post-saddle stage, Hofman et al. [165] concentrated on a $^{240}$Cf compound nucleus for which the saddle-to-scission path dominates. In contrast to all previous works which support an increase of $\beta$ with deformation, they concluded that $\beta^{post-sad}$ is smaller than $\beta^{pre-sad}$. This small post-saddle contribution is related to the assumption made in Ref. [165] about the *a priori* large pre-saddle value $\beta^{pre-sad} \approx 20\cdot10^{21}$s$^{-1}$ from [164]. As mentioned above, the correlation between the pre- and post-saddle contributions implies that a large $\beta^{pre-sad}$ value requires a small $\beta^{post-sad}$ strength, and vice versa, to agree with the experimental data. A re-analysis of the $^{240}$Cf measurement was done by Shaw et al. [166] who achieved an equally good description of the data with $\beta^{pre-sad} \approx 4\cdot10^{21}$s$^{-1}$ combined with $\beta^{post-sad} \approx 20\cdot10^{21}$s$^{-1}$ provided that the temperature-dependence of the level density is properly taken into account. It should be pointed out that in most of the above cited works the fission-decay width $\Gamma_f(t)$ was described with an exponential in-growth function [163]. It was demonstrated [122] that this parameterisation deviates from the numerical solution and leads to an overestimation of $\beta$. This contributes to a large part to the discrepancy in the extracted values for $\beta$.

In contrast to the controversial situation outlined above that exists at low and moderate CN excitation energies, a consensus might be expected to emerge at high excitation energies where transient effects should manifest more strongly. A few data on fusion-induced fission exist at higher excitation energies. In Ref. [167] for symmetrically fissioning Pt nuclei at $E^* \approx$ 500 MeV, a total pre-scission delay of $\tau_{trans}+\tau_{ss} \approx 8 \cdot 10^{-21}$s was extracted. Mordhorst *et al.* [168] investigated compound nuclei between Po and Am with $E^*$ in the range of (200-570) MeV. For symmetric mass splits, they estimated the sum $\tau_{trans}+\tau_{ss}$ to be around $(20-30) \cdot 10^{-21}$s with a dominating contribution from $\tau_{ss}$. The incomplete fusion mechanism exploited in Ref. [168] required making cuts in the linear momentum transfer distribution. As shown in Ref. [169], gates on the recoil velocity can lead to spurious results if non-representative samples of events are selected.

In Refs. [157] and [158], model calculations accounting for the combined influence of $\tau_{trans}$, $K$ and $\tau_{ss}$ were performed to explain measured pre-scission neutron multiplicities. Special care was taken to include transient effects in a way similar to the present work. For the $^{158}$Er compound nucleus excited to energies around 150 MeV, Grangé *et al.* [157] extracted an upper limit for the dissipation strength of about $5 \cdot 10^{21}$s$^{-1}$, while Gavron *et al.* [158] arrived at $\beta \approx 6 \cdot 10^{21}$s$^{-1}$ and $\tau_{trans} \approx (3-8) \cdot 10^{-21}$s. It is worth pointing that our conclusion for $\tau_{trans}$ is in rather good agreement with Refs. [157, 158] although different signatures are employed. The temperature-dependence of the transient time derived by microscopic calculations [9] is able to explain quantitatively the remaining difference between the value of $\tau_{trans}$ extracted in Refs. [157, 158] and in this work, which, on average, is concerned with higher excitation energies than Refs. [157, 158]. The agreement obtained for $\tau_{trans}$ constrains $\beta$ to values around $5 \cdot 10^{21}$s$^{-1}$ at small deformation. Lighter compound nuclei are expected to be characterized by more elongated saddle-point shapes than the heavy systems considered here. This is partly compensated by the higher angular momentum induced in the fusion reaction, which results in a more compact saddle-point configuration of the $^{158}$Er compound nucleus. Therefore, only a limited range of deformations is probed. Combining the present result, which is exclusively sensitive to the pre-saddle region, with the conclusion of Refs. [157, 158], which is based on the entire path up to scission, gives insight into dissipation at large deformation. The good description achieved with a constant $\beta$ in Refs. [157, 158] and the value obtained here for $\beta^{pre-sad}$ suggests a nearly deformation-independent dissipation strength up to the largest elongations experienced along the decay of the $^{158}$Er nucleus. The present comparison seems therefore to exclude a strong deformation-dependence of $\beta$. Dynamical calculations [159, 170] which succeeded in describing a wide range of experimental data of heavy nuclei support this point of view.

**b. Transfer-induced fission**

Fission time scales were also extracted from experiments using multi-nucleon transfer reactions in normal kinematics for inducing fission, see e.g. [171, 172]. The detection of the projectile-like fragment allowed deducing the mean excitation energy, angular momentum and size of the excited fissile target-like nucleus. The analysis of the measured fission excitation functions $P_f(E^*)$ led to a transient time in the range $(5-20) \cdot 10^{-21}$s for 50 MeV $< E^* <$ 75 MeV for actinide compound nuclei [171]. On the basis of predictions for $\tau_{ss}$ [173] it was concluded that $\tau_{trans}$ dominates the dynamical time before scission. The discrepancy between the transient time value of [171] and the present one may be due to several reasons. The lower temperatures involved in transfer-induced fission is expected to lead to larger $\tau_{trans}$, whereas the deformation of the actinide target is conjectured to decrease $\tau_{trans}$. In addition, the omission of Kramers' factor in the calculations used in some of these works may artificially increase the extracted value of $\tau_{trans}$.

### c. Spallation-induced fission

#### α) Experiments in normal kinematics

In most experiments exploiting the spallation reaction mechanism in normal kinematics to induce fission, see Refs. [77, 78, 22, 174] and references therein, the number of light particles emitted during the reaction was used to extract the initial excitation energy of the system on an event-by-event basis. The analysis of the $P_f(E^*)$ data yielded a very small value with an upper limit of $(0.5-1) \cdot 10^{-21}$s in Refs. [77, 78] or even zero in Ref. [22] for $\tau_{trans}$. Similarly, neutron multiplicities were described considering a finite delay only for the descent from the saddle to the scission point [175]. The conclusion of (nearly) inexistent early transient effects in the quoted articles is based on calculations using a level density $a_f/a_n$ ratio equal to 1. For heavy systems, there are theoretical arguments [124, 125] for surface effects leading to $a_f/a_n > 1$. To illustrate the influence of the modelling of the level densities, different observables are presented in Fig.15. In the left panel, the experimental evaporation residue cross section, the complement of $P_f(E^*)$, measured in a spallation reaction at GSI [176] is compared with two different calculations. The result obtained with $a_f/a_n = 1 \oplus \tau_{trans} \approx 0$s as used, e.g. in Ref. [22], matches quite well the cross sections computed with $a_f/a_n > 1 \oplus \tau_{trans} \neq 0$s, and is in gross agreement with the experiment. A finite transient delay decreases the fission probability whereas $a_f/a_n > 1$ increases it [76, 49]. Another independent illustration of possible misinterpretation related to model parameters if only cross section measurements are available can be found in [177]. The $\sigma_Z$ signature allows going a step further [69]. As seen in the right panel of Fig.15, the measured Z-widths are inconsistent with the calculation omitting transient effects. This is understood by the fact that $\sigma_Z$ depends principally on $T^{sad}$ while observables connected to the fission probability are strongly affected by additional critical parameters (see also the discussion presented in Refs. [51, 52]). Its weaker model-dependence is a significant advantage of the $\sigma_Z$ signature.

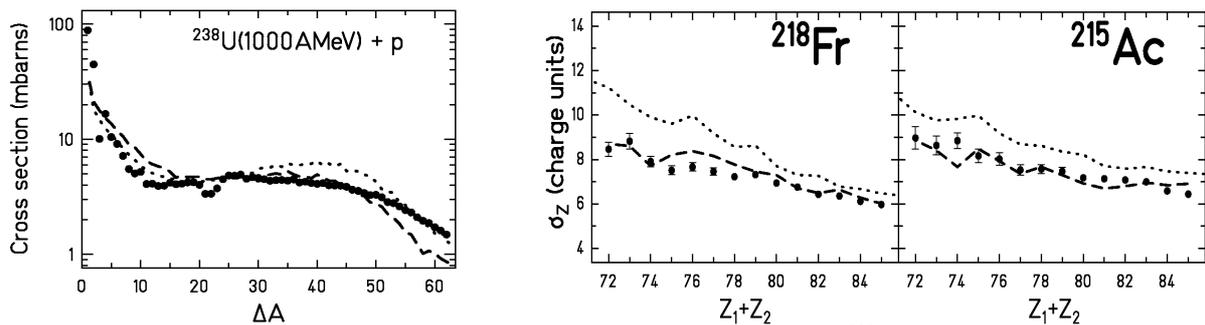

**Figure 15:** Left: Evaporation residue cross section for the spallation reaction $^{238}$U (1000 $A$MeV) + $p$ [176] as a function of mass loss with respect to the projectile mass. Right: Width $\sigma_Z$ as a function of $Z_1+Z_2$ as measured in the present experiment for two different beams. In all panels, the data (dots) are compared with calculations performed for ($a_f/a_n \neq 1 \oplus \tau_{trans} \neq 0$s) (dashed lines) and calculations done with ($a_f/a_n = 1 \oplus \tau_{trans} = 0$s) (dotted lines).

A successful description of experimental excitation functions without introducing any transient delay suggests that "fission is decided upon very fast and early in the long de-excitation chain", according to Ref. [22]. The authors emphasize the minimal collective distortion induced in a spallation reaction. In our understanding, that is equivalent to the statement that the nucleus is initially characterized by a shape resembling the ground-state configuration of the heavy reaction partner. According to the dynamical picture of fission based on the solution of the equation of motion [2, 8] a finite time is required for evolving from the spherical ground-state configuration up to the saddle point. The conclusion of a transitional saddle state which is immediately populated ($\tau_{trans} \approx 0$s) suggests that the spreading of the collective degrees of freedom in the fission direction is much faster than expected from dynamical calculations. This questions some of the basis of nuclear dynamics which seemed well-established.

**β) Experiments in inverse kinematics**

The research program on spallation in inverse kinematics, conducted at GSI, gave rise to a large collection of data which can be used to benchmark theoretical predictions. Up to now, various beams between $^{56}$Fe and $^{238}$U at energies from 150 *A*MeV up to 1500 *A*MeV have been used to irradiate proton and deuterium targets [178]. Though motivated by technical applications [179, 180], these data contain rich fundamental physics information. In particular, the heaviest systems were used to investigate fission dynamics. In the pioneering work on $^{197}$Au (800 *A*MeV) + *p* [75], a transient time of $(3\pm1)\cdot10^{-21}$s was extracted, in agreement with our result. The estimated dissipation strength $\beta \approx (2-3)\cdot10^{21}$s$^{-1}$ is lower than the one obtained in the present work. The reason for the discrepancy is currently not clear. Initial compound nucleus deformation effects are not expected to play a role according to the spherical character of $^{197}$Au. The conclusions of Ref. [75] rely on the total fission cross section, which might be insufficient to unambiguously determine $\beta$ as discussed above and shown in Ref. [133]. The authors note that any observable strongly related to $T^{sad}$ would be more relevant. The analysis of cross sections measured in spallation experiments with a $^{238}$U beam at 1000 *A*MeV [76, 181] yielded a dissipation strength around $2\cdot10^{21}$s$^{-1}$ as well. As discussed below, deformation effects are conjectured [73] to lead to an artificially small dissipation strength when transient effects are modelled according to Refs. [121, 122] which was done in Refs. [76, 181] for initially deformed systems.

**d. Fission induced by fragmentation of stable heavy ions**

As mentioned already, fission and evaporation residue cross sections from the interaction of $^{208}$Pb and $^{238}$U projectiles at relativistic energies on various targets [74] did not allow an unambiguous determination of $\beta$. The pioneering work of Jurado *et al*. [68] introduced the Z-width signature for the fragmentation-induced fission of a $^{238}$U beam at 1000 *A*MeV on a plastic target. A transient delay of $\tau_{trans} \approx (1.7\pm0.4)\cdot10^{-21}$s was extracted. This value was related to $\beta \approx 2\cdot10^{21}$s$^{-1}$ according to the $\Gamma_f(t)$ formula derived for initially spherical nuclei [121, 122]. The latter parameterisation is inappropriate for describing the decay of the deformed pre-fragments produced in the fragmentation of the $^{238}$U projectile [35, 182]. Its use might biases the value extracted for $\beta$, and lead to an underestimation of the dissipation strength [73]. However, the value deduced in Ref. [68] for $\tau_{trans}$ from the Z-width observable is not, or only weakly, affected because the fission delay governs the temperature at saddle directly. The transient delay of [68] was found to be consistent with the present conclusion once the expected reduction of the transient time due to initial deformation effects is included [73, 182].

**3) Dependence on temperature**

The data set collected during the present experiment is well described under the assumption of a constant dissipation strength for the deformation path from a spherical configuration up to the saddle point, and a transient time which is independent of the temperature of the compound nucleus in the range of 2.4 MeV < $T$ < 5.5 MeV. As discussed above, according to the experimental uncertainties, the present conclusion on $\tau_{trans}$ is consistent with the calculations of Sargsyan *et al.* [9] who predict only a moderate decrease of $\tau_{trans}$ within this $T$ range for $\beta = 4.5 \cdot 10^{21}$s$^{-1}$. According to these calculations, a difference between the data of the present work, on average concerned with data at $T > 3$ MeV, and measurements done below $T = 3$ MeV is expected. This is where the variation of $\tau_{trans}$ with $T$ is predicted to be strong. This conjecture seems to be corroborated by the discussion in the previous section. Indeed, our observations of the gross $T$-dependence of $\tau_{trans}$ above follow the theoretically predicted tendency around $\beta \approx 5 \cdot 10^{21}$s$^{-1}$.

The temperature-dependence of nuclear dissipation is among the most controversial points in nuclear dynamics since years. As mentioned above, Hofman *et al.* [134] introduced a strong rise of $\beta$ with temperature in the range 1.2 MeV < $T$ < 1.8 MeV. Several studies confirmed the pertinence of this ansatz, while many others did not require it, and the question remains pending (see previous section as well as Refs. [41, 52] and references therein). The extrapolation of the prescription of Hofman *et al.* to the temperature domain of the present experiment does not reproduce the measured $Z$-widths. It is also inconsistent with the predictions of Sargsyan *et al.* [9]. Recently, MacCalla and Lestone [51] claimed that the $T$-dependence derived in Ref. [134] "is an artefact generated by an inadequate fission model". They disagree with the modelling of the level density in standard statistical model calculations, and demonstrated that omitting the temperature-dependence of the location of the saddle point leads to an overestimation of $a_f$ at high excitation energies. This can be compensated by an artificially increasing of $\beta$ with $T$. Within a deterministic model, Wilczynski *et al.* [183] used the number of neutrons emitted along quasi-fission trajectories at 2.2 MeV $\leq T \leq$ 2.4 MeV as a "friction meter". Their study led to a rather small dissipation coefficient for the entrance stage, and a very large strength for the phase of re-separation of the system into two fragments. Combined with the $T$-dependence derived in Ref. [134], Wilczynski *et al.* interpreted their result as the modification of the nature of dissipation (from one- [16] to two- [17] body kind) at high temperature. A similar investigation of the quasi-fission mechanism [184] did not confirm the conclusion of Ref. [183] for $T \approx 2$ MeV. Fission-fragment mass and angular distributions were found to be consistent with the one-body dissipation theory. Assuming a temperature-independent dissipation strength of one-body nature for $T > 2.4$ MeV as derived in the present work, the conclusion of Ref. [183] implies a very abrupt change of the viscous nature of nuclear matter with intrinsic excitation. Further studies are clearly necessary to investigate this point.

## VII. Summary and conclusions

An innovative experimental approach for probing transient effects in nuclear fission, and more generally, the magnitude of nuclear dissipation at small deformation, is reported. For the first time, the de-excitation of *highly excited and highly fissile initially spherical nuclei at low angular momenta* was carefully studied, thereby allowing an unambiguous test of how viscosity influences the establishment of quasi-equilibrium in nuclear matter. The experiment was realized under very specific and favourable initial conditions, involving the state-of-the-art installations at GSI, Darmstadt. Projectile-fragmentation was used in a two-step reaction scenario. The identification of the products resulting from the fragmentation of a primary stable $^{238}$U beam by the GSI fragment separator allowed to prepare beams of radioactive fissile spherical nuclei which fragment again after interaction in a secondary target. This leads to the formation of highly excited and fissile (nearly) spherical compound nuclei. For those secondary fragmentation compounds which decay by fission, the fission-fragment nuclear-charge distributions are accurately measured. The two fragments are detected in coincidence, which permits to classify the data according to initial excitation energy and fissility. The width $\sigma_Z$ of the distribution is used as a sensitive probe of pre-saddle relaxation effects. The analysis, within the framework of a time-dependent statistical model, strongly supports the manifestation of transient effects which delay the establishment of a quasi-stationary flow across the fission barrier at high excitation energies. For initially spherical compound nuclei, an average transient time of $\tau_{trans} \approx (3.3 \pm 0.7) \cdot 10^{-21}$s is extracted. That implies a dissipation strength of $\beta = (4.5 \pm 0.5) \cdot 10^{21} \text{s}^{-1}$ at small deformations. The classification of the data according to initial excitation energy and fissility suggests that neither the fission delay nor the dissipation strength does strongly depend on temperature and fissility for 2.4 MeV < $T$ < 5.5MeV and 30.5 < $Z^2/A$ < 37.5, respectively. The size of the data set, including the fragmentation-induced fission of 45 radioactive actinide beams *measured and analyzed under identical conditions*, further supports the extracted conclusions.

The present result, which is based on well-defined initial conditions, on a specifically sensitive signature and on elaborate model calculations, gives new insight into the controversy that exists on the very existence of transient effects and on the nature and magnitude of dissipation in nuclear matter. The discussion of a wide compilation of previous observations suggests that some inconsistent interpretations might originate, at least partly, from different initial conditions. The results of the present work, which is restricted to the pre-saddle regime, can be combined with conclusions based on data integrating the entire fission path up to the scission point, suggesting that the deformation-dependence of $\beta$ is rather weak.

The conclusion on the magnitude of $\beta$ and its temperature dependence extracted in the present work is in favour of an over-damped motion of one-body nature at small deformation and high excitation energy. The extracted dissipation strength and transient time are, nonetheless, reduced compared to the predictions of the early one-body theory for compact shapes [16]. In a study of an extended set of observables, which are to a great part sensitive to the whole fission path, multi-dimensional dynamical calculations [25, 159, 185, 186] arrived at a similar conclusion for both, the viscous nature of nuclear matter, and the value of $\beta$. More generally, theoretical studies on surface motion of a cavity wall [18, 187, 188, 189], among which multipole vibrations and fission are, suggest that the original wall damping mechanism [16] overestimates the dissipation rate in nuclear matter. That is understood by chaotic single-particle motion and quantal effects, which violate the original assumption of full randomisation [190]. This deficiency in the early derivation of the one-body theory partly contributes to the variety of contradicting conclusions on nuclear dissipation [191]. The accurate numerical value extracted in this work for $\beta$ needs, nevertheless, to be considered with caution, since it is based on a one-dimensional equation of motion. Realistic simulations [192] require a three-dimensional deformation space. Yet, the close agreement with the above quoted multi-dimensional calculations is worth noting. The transient time, which is rather directly reflected in the experimental $\sigma_Z$ signature, should not be affected by this limitation. Very recently, Sadhukhan and Pal [193] discussed the accuracy of extracting the strength of "pres-saddle dissipation" using Kramers decay-width in the statistical model. The stochastic nature of fission implies that the so-extracted strength for the pre-saddle region is influenced by the post-saddle dynamics due to backstreaming trajectories which experience also dissipation beyond saddle. It would be of interest to quantify the influence of this post-saddle contribution to the Z-width with a realistic deformation-dependent dissipation within a three-dimensional Langevin approach [192]. We believe that it is small as compared to the what derived in [193] using a schematic deformation-dependent $\beta$ in a one-dimensional Langevin equation and less sensitive signatures. Note also that a large part of the pre-saddle time is spent oscillating about the equilibrium shape, yielding more weight to the pre-saddle contribution. The transient time extracted in the present work is expected to be essentially unaffected by this aspect of the dynamics.

The successful reproduction of the entire data set together with previous results of dynamical calculations up to scission [159, 170, 185, 186, 187] indicates that diffusion and friction inside and outside the quasi-bound region are consistently described by the conventional Fokker-Planck or Langevin dynamics with a rather universal dissipation strength. No departure from normal diffusion dynamics is observed. According to Einstein's relation, the nearly constant $\beta$ suggests that diffusion scales with temperature [19].

The present work demonstrates that peripheral heavy-ion collisions at relativistic energies constitute a powerful approach for the investigation of fission dynamics over a wide domain of temperatures. The specific features of fragmentation-induced processes permit tailoring the initial conditions, in terms of deformation, excitation energy and angular momentum of the compound nucleus. Complicating effects which hinder the extraction of reliable transient time and dissipation strength values are avoided in the analysis due to high excitation energies and small angular momenta of the fissioning system. In addition, the availability of a two-step fragmentation reaction scenario together with a pertinent choice of the secondary projectiles allows to restrict the data set to initially nearly spherical fissile systems, which avoids the influence of initial deformation. To deepen our understanding of the temperature- and fissility-dependences, exclusive measurements are required for an accurate characterisation of the decaying system. Such studies are foreseen at the future FAIR facility within the Reactions with Relativistic Radioactive Beams ($R^3B$) project [194]. A set-up which allows the identification of both fission fragments in nuclear mass, nuclear charge and energy combined with light-particle and $\gamma$-ray detectors will provide a complete reconstruction of the reaction kinematics. The studies of fission dynamics by means of peripheral heavy-ion collisions at relativistic energy merit to be pursued in parallel to conventional methods. The combination of information derived from different approaches is the only way to unravel various contributions to the puzzling fission process.

**Acknowledgements**

Two of us (C.S. and P.N.N.) are grateful for the post-doctoral position at GSI granted by the A. von Humboldt foundation. This work has been financially supported in part under U.S. DOE Grant Number DE-FG02-91ER-40609 and by the French-German GSI/IN2P3/CEA Contract Number 04-48.

**Appendix I: Comparison of $\sigma_Z$ data as obtained with different techniques**

Most experimental data on fission consists of cross sections, fission-fragment mass and kinetic energy distributions collected from nuclear collisions with beam energies below 10 $A$MeV. Information on the element number $Z$ of the fragments remains scarce, in particular for the heavy fragment, see e.g. [195, 196, 197]. In the peculiar case of low-energy fission, measuring the $Z$ of the light fragment allows to extract the nuclear charge of the heavy partner since light-charged particle evaporation of the latter is unlikely. Radiochemical and $\gamma$-spectroscopy [198, 199, 200, 201] methods have been employed to determine $Z$ of the heavy reaction product. Unfortunately, these techniques do not allow systematic investigations and the information is restricted to the isotopes of a few elements. Recently, a large campaign, which the present experiment is part of, was initiated at GSI to measure the full nuclide production in nucleus-nucleus collisions at relativistic energies, see Refs. [94, 202] and references therein. These experiments utilized inverse kinematics which allows the precise determination of atomic numbers of all fission fragments up to the heaviest ones. In the following, we compare the present results for $\sigma_Z$ to a compilation of representative previous data.

### a. Fusion-induced fission

Heavy-ion fusion-induced fission provided a wealth of information on experimental fission-fragment mass distributions and widths $\sigma_A^{exp}$. To compare the latter with the $\sigma_Z^{exp}$ reported in this work, we suggest to convert mass widths into nuclear-charge widths by applying the unchanged charge density (UCD) hypothesis, i.e. $\sigma_Z^{UCD} = (Z_{CN}/A_{CN}) \cdot \sigma_A^{exp}$ where $Z_{CN}$ ($A_{CN}$) corresponds to the nuclear charge (mass) of the CN formed in the fusion reaction.

In a fusion-induced fission reaction, $Z_{fiss}$ is quite well-defined and very close to $Z_{CN}$, whereas $Z_{fiss}$ covers a broad distribution in fragmentation-induced fission. We have shown (see section IV) that $Z_1+Z_2$ is a well-suited observable for selecting the parent element in the latter reaction scenario. In most fusion-induced fission data, $E^*$ ranges from 30 up to 150 MeV. At the lower excitation energy limit, first-chance fission and neutron evaporation dominate, while at higher $E^*$ up to two light-charged particles can be emitted before scission [203, 204, 78]. Therefore, to make the comparison with the fusion-induced data meaningful, the $\sigma_Z^{exp}$ measured in this work are filtered with respect to $Z_1+Z_2 = Z_{CN}$, ($Z_{CN}$ -1) and ($Z_{CN}$ - 2) corresponding, respectively, to a mean initial excitation energy $E^*$ of 25 MeV, 70 MeV and 130 MeV. The results of such a procedure are summarized in Table A.1. Two values for the initial CN excitation energy are given for fusion experiments: $E^*$ and $E^*(l)$ are, respectively, the intrinsic excitation energy without and with taking the rotational energy into account [60].

| CN | $E^*$ (MeV) | $E^*(l)$ (MeV) | $\sigma_A^{exp}$ | $\sigma_Z^{UCD}$ | $\sigma_Z^{exp}$ for $Z_1+Z_2=Z_{CN}$ | $\sigma_Z^{exp}$ for $Z_1+Z_2=Z_{CN}-1$ | $\sigma_Z^{exp}$ for $Z_1+Z_2=Z_{CN}-2$ |
|---|---|---|---|---|---|---|---|
| $^{205}$At | 57.9<br>67.6<br>74.0<br>98.2<br>108.7<br>138.4 | 49.6<br>56.2<br>60.4<br>76.8<br>81.5<br>104.8 | 13.60<br>14.25<br>15.59<br>19.00<br>19.72<br>23.26 | 5.64<br>5.91<br>6.46<br>7.88<br>8.18<br>9.64 | 5.29 | 5.33 | 5.81 |
| $^{206}$At | 76.2<br>150.6 | 65.4<br>131.6 | 12.33<br>16.28 | 5.09<br>6.72 | 5.18 | 5.47 | 5.68 |
| $^{209}$At | 73.2 | 58.5 | 13.56 | 5.52 | - | - | - |
| $^{209}$Rn | - | - | - | - | 5.18 | 5.45 | 5.80 |
| $^{214}$Rn | 88.1 | 74.1 | 14.97 | 6.02 | - | - | - |
| $^{212}$Fr | - | - | - | - | 5.25 | 5.41 | 5.89 |
| $^{213}$Fr | 86.0 | 71.5 | 16.00 | 6.54 | - | - | - |
| $^{215}$Fr | 112.9 | 95.7 | 16.49 | 6.67 | - | - | - |
| $^{217}$Fr | - | - | - | - | 5.55 | 5.62 | 5.88 |
| $^{218}$Ra | 61.2 | 53.3 | 14.25 | 5.75 | 5.40 | 5.56 | 5.87 |
| $^{217}$Ac | 92.8 | 80.0 | 18.65 | 7.65 | 4.87 | 5.15 | 5.70 |
| $^{220}$Th | 55.6 | 47.4 | 15.36 | 6.28 | - | - | - |
| $^{221}$Th | - | - | - | - | 4.82 | 5.11 | 5.65 |
| $^{222}$Th | 73.0 | 66.5 | 17.20 | 6.97 | 4.91 | 5.35 | 5.78 |
| $^{224}$Th | 53.8 | 49.8 | 14.97 | 6.01 | 5.60 | 6.02 | 6.36 |
| $^{225}$Pa | 30 – 50[a] <br> 32 – 44[b] | - <br> - | 16.43 - 17.03 <br> 14.10 – 15.60 | 6.64 – 6.89 <br> 5.70 – 6.31 | - | - | - |
| $^{226}$Pa | - | - | - | - | 6.41 | 6.90 | 7.00 |
| $^{227}$Pa | 30 – 32[b] | - | 15.60 – 14.30 | 6.25 – 5.73 | 6.90 | 7.32 | 7.32 |
| $^{228}$U | 37 – 50[a] <br> 35.11 – 46.34[c] | - | 17.03 – 17.89 <br> 13.00 – 15.70 | 6.87 – 7.22 <br> 5.24 – 6.34 | - | - | - |
| $^{231}$U | - | - | - | - | 7.83 | 9.02 | 8.75 |
| $^{229}$Np | 72.9 | - | 17.44 | 7.08 | - | - | - |

**Table A.1** : Four first columns: characteristics of the fusion-fission experiments. The data are taken from [60, 61] and references therein, except [a] from [205], [b] from [198] and [c] from [199]. Fifth column: estimate of the Z-width based on the UCD assumption and on the measured mass width of the fourth column, see the text. Three last columns: presently measured Z-widths gated by $Z_1+Z_2 = Z_{CN}$, ($Z_{CN} - 1$) and ($Z_{CN} - 2$).

As seen in the table, for the compound nuclei $^{228}$U and $^{225}$Pa, in spite of similar entrance channels, the authors of Refs. [198, 199] and [205] announce different values for $\sigma_A^{exp}$. The former studies employed radiochemical method which might miss some elements, rendering the mass distribution narrower. Taking this into account, the overall comparison shows that $\sigma_Z^{exp}$ is slightly smaller than $\sigma_Z^{UCD}$. The difference may be understood on the basis of temperature and angular momentum effects. From the parameterisation of $d\sigma_A^2/dL^2$ established in [60], the influence of $L^2$ is of the order of a few percents for $32 < Z^2/A < 36$. Assuming, that on average, the induced angular momentum is about 10 ℏ for relativistic heavy-ion collisions [87] and $L \approx 30$ ℏ for fusion reactions [60, 61], one might expect larger Z-widths by about 0.5 charge unit in fusion-induced fission. Thais matches reasonably well with the numbers given in Table A.1, at least, at excitation energies which are not too high. The increase of the discrepancy with increasing $E^*$ might be caused by the increasing importance of the CN temperature on the Z-width. Finally, some differences may be explained by charge polarisation [106] (which violates the UCD assumption) and contribution from quasi-fission [206] (which leads to broader mass distributions) in the fusion-induced data.

**b. Spallation-induced reactions**

The strategy adopted so far in most spallation experiments at GSI differs from the present one because a large part of the physics program was concerned with non-fissioning evaporation residue events. Here, a stable beam impinged on a target of liquid hydrogen or deuterium. Due to the limited acceptance of the FRS, only one of the fission fragments was transmitted and subsequently identified in mass and nuclear charge. The FRS also provided a very precise velocity measurement, which does retain some information on the unobserved fission partner [101]. However, the precise value of $Z_1+Z_2$ is unknown on an event-by-event basis. Only the mean $<\sigma_Z>$, averaged over the whole $Z_1+Z_2$ range, is available.

Some of the results obtained at GSI in spallation experiments with relativistic heavy-ion beams in inverse kinematics are summarized in Table A.2. The larger widths for deuteron as compared to hydrogen targets are due to the higher $E^*$ involved in the former case [101]. The influence of the difference in beam energy when comparing the fragmentation and spallation data considered here is negligible. Nonetheless, relativistic collisions between two heavy ions lead to pre-fragments which populate higher $E^*$ than light particle-induced reactions do, see e.g. [83, 85]. Furthermore, angular momenta up to about 50 ℏ can be reached in spallation [86] while $L$ remains below 20 ℏ for fragmentation [87]. The influence of temperature and angular momentum on the average $Z$-width might thus partly compensate each other, leaving similar values for the two mechanisms. The $<\sigma_Z>$ measured in the present work amount to (6-7) charge units for the lightest beams and up to 8 units for the Pa and U projectiles, in gross agreement with the values of Table A.2.

| Reaction | $E_{proj}$ (AMeV) | $\sigma_A^{exp}$ | $\sigma_Z^{exp}$ |
|---|---|---|---|
| $^{197}$Au + p [75] | 800 | 15.1±2.0 | 6.2±1.0 |
| $^{208}$Pb + p [108] | 500 | 15.1±0.6 | 6.3±0.2 |
| $^{208}$Pb + p [101] | 1000 | 16.1±0.8 | 6.6±0.3 |
| $^{208}$Pb + d [101] | 1000 | 17.4±1.0 | 7.3±0.5 |
| $^{238}$U + p [106] | 1000 | 17.5±0.5 | 7.0±0.2 |
| $^{238}$U + d [107] | 1000 | 20.0±0.5 | 7.7±0.2 |

**Table A.2:** Experimental results from spallation-induced fission data measured at GSI.

### c. Peripheral heavy-ion collisions with relativistic stable beams

The interaction between $^{238}$U projectiles at 750 AMeV on lead [207], for which only one of the fragments was detected, were measured to have a $<\sigma_Z>$ of (6.9±0.4) charge units, in agreement with the values obtained in the present work. Recently [68], with a set-up similar to the one of Fig.1, bottom panel, the interaction of $^{238}$U at 1000 AMeV on a plastic target was investigated. Neither the difference in the incident energy, nor in the target material [208, 209] is expected to lead to a noticeable difference compared with the present measurement. The evolution of $\sigma_Z$ with $Z_1+Z_2$ for the $^{238}$U beam ($Z_{proj}^2/A_{proj} = 35.56$) is superimposed to the results obtained for the $^{231-234}$U projectiles ($36.33 < Z_{proj}^2/A_{proj} < 36.64$) of this work in Fig.A.1. Within the error bars, the two data sets agree with each other. The slightly larger widths obtained with the radioactive beams could very tentatively be assigned to a fissility effect. Nonetheless, projectiles differing in mass imply slightly different $E^*$ for a given $Z_1+Z_2$. Dedicated exclusive experiments are required to settle this point.

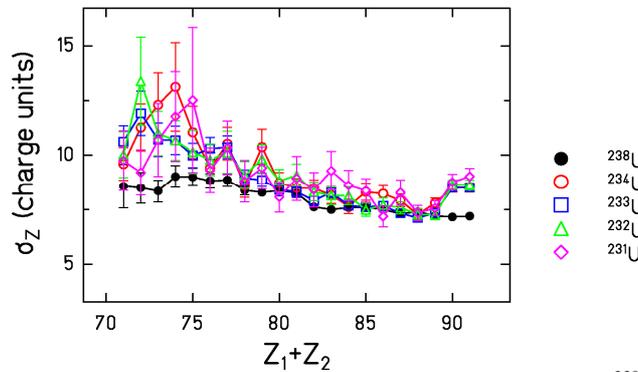

**Figure A.1:** Width $\sigma_Z$ as a function of $Z_1+Z_2$ for fragmentation-induced fission of $^{238}$U at 1000 AMeV on plastic (dots) from [68] and for $^{231-234}$U at ≈ 420 AMeV from this work (symbols as indicated).

Summarizing, within the experimental uncertainties and according to the properties inherent to different entrance channels, our data agree with those obtained in the past *via* various techniques.